\documentclass[sigconf]{acmart}

\usepackage{booktabs} 
\usepackage{balance}

\def\BibTeX{{\rm B\kern-.05em{\sc i\kern-.025em b}\kern-.08emT\kern-.1667em\lower.7ex\hbox{E}\kern-.125emX}}

\usepackage{amsfonts}
\usepackage{subcaption}
\usepackage{float}
\usepackage{multirow}

\copyrightyear{2021}
\acmYear{2021}
\setcopyright{iw3c2w3}
\acmConference[WWW '21]{Proceedings of the Web Conference 2021}{April 19--23, 2021}{Ljubljana, Slovenia}
\acmBooktitle{Proceedings of the Web Conference 2021 (WWW '21), April 19--23, 2021, Ljubljana, Slovenia}
\acmPrice{}
\acmDOI{10.1145/3442381.3450014}
\acmISBN{978-1-4503-8312-7/21/04}

\author{Christian Hansen}
\authornote{Both authors share the first authorship.}
\affiliation{
  \city{University of Copenhagen}
}
\email{chrh@di.ku.dk}

\author{Casper Hansen}
\authornotemark[1]
\affiliation{
  \city{University of Copenhagen}
}
\email{c.hansen@di.ku.dk}

\author{Jakob Grue Simonsen}
\affiliation{
  \city{University of Copenhagen}
}
\email{simonsen@di.ku.dk}

\author{Stephen Alstrup}
\affiliation{
  \city{University of Copenhagen}
}
\email{s.alstrup@di.ku.dk}

\author{Christina Lioma}
\affiliation{
  \city{University of Copenhagen}
}
\email{c.lioma@di.ku.dk}

\begin{document}
\fancyhead{}


\begin{abstract}
Semantic hashing represents documents as compact binary vectors (hash codes) and allows both efficient and effective similarity search in large-scale information retrieval. The state of the art has primarily focused on learning hash codes that improve similarity search effectiveness, while assuming a brute-force linear scan strategy for searching over all the hash codes, even though much faster alternatives exist. One such alternative is multi-index hashing, an approach that constructs a smaller candidate set to search over, which depending on the distribution of the hash codes can lead to sub-linear search time.
In this work, we propose Multi-Index Semantic Hashing (MISH), an unsupervised hashing model that learns hash codes that are both effective and highly efficient by being optimized for multi-index hashing. We derive novel training objectives, which enable to learn hash codes that reduce the candidate sets produced by multi-index hashing, while being end-to-end trainable. In fact, our proposed training objectives are model agnostic, i.e., not tied to how the hash codes are generated specifically in MISH, and are straight-forward to include in existing and future semantic hashing models.
We experimentally compare MISH to state-of-the-art semantic hashing baselines in the task of document similarity search. We find that even though multi-index hashing also improves the efficiency of the baselines compared to a linear scan, they are still upwards of 33\% slower than MISH, while MISH is still able to obtain state-of-the-art effectiveness.

\end{abstract}


\title{Unsupervised Multi-Index Semantic Hashing}
\maketitle

\section{Introduction}

Similarity search is a fundamental information retrieval task that aims at finding items similar to a given query. Efficient and effective similarity search is essential for a multitude of retrieval tasks such as collaborative filtering, content-based retrieval, and document search \cite{wang2018survey,wang2015learning}. Semantic Hashing \cite{salakhutdinov2009semantic} methods enable very efficient search by learning to represent documents (or other types of data objects) as compact bit vectors called \emph{hash codes}, where the Hamming distance is used as the distance metric between hash codes. In this setting, similarity search is expressed as either radius search (finding all hash codes with a specified maximum Hamming distance), or as k-nearest neighbour (kNN) search by incrementally increasing the search radius until the Hamming distance to the $\text{k}^{\text{th}}$ document is equal to the search radius. 
Early work on semantic hashing \cite{weiss2009spectral,zhang2010laplacian} was inspired by techniques similar to spectral clustering \cite{ng2002spectral} and latent semantic indexing \cite{deerwester1990indexing}, whereas modern approaches use deep learning techniques, typically unsupervised autoencoder architectures where hash codes are optimized by learning to reconstruct their original document representations \cite{chaidaroon2017variational,chaidaroon2018deep,nash2018,hansensemhash2019,dong-etal-2019-document,hansen2020PairRec}. This line of work has led to extensive improvements of the effectiveness of document similarity search, but has had a lesser focus on efficiency, as it uses a brute-force linear scan of all the hash codes. While the highly efficient Hamming distance does enable large-scale search using linear scans \cite{shan2018recurrent}, significantly faster alternatives exist. Multi-index hashing \cite{norouzi2012fast,norouzi2012fast-initial-paper,greene1994multi} is such an alternative, which, depending on a query hash code, can enable sub-linear search time by constructing a smaller set of candidate hash codes to search over. Hash codes can be used extremely efficiently as direct indices into a hash table for finding exact matches, however when doing radius search the number of such hash table lookups grows exponentially. Multi-index hashing is based on the observation that by splitting the hash codes into $m$ substrings and building a hash table per substring, the exponential growth of the number of lookups can be significantly reduced. In practice, the efficiency of multi-index hashing is heavily dependent on the distribution of the hash codes, and most particularly their substrings, which affects the size of the constructed candidate set for a given query hash code. However, no existing semantic hashing methods consider this aspect, which we experimentally verify limits their efficiency.

\begin{figure}
    \centering
    \includegraphics[width=0.95\linewidth]{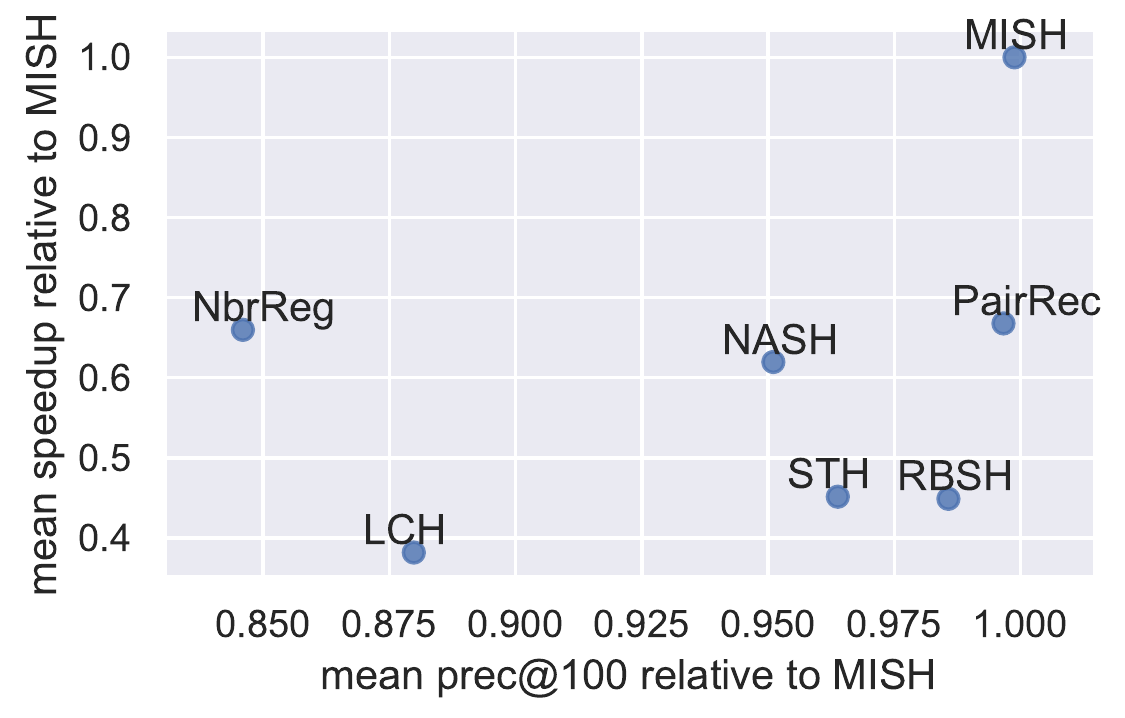}
    \vspace{-13pt}
    \caption{Method comparison relative to our proposed MISH, averaged over all used datasets. We plot each method as a point regarding its mean speedup of multi-index hashing compared to a linear scan, as well as mean prec@100, relative to MISH.\label{fig:eeplot}}
    \vspace{-13pt}
\end{figure}

To address the above efficiency problem, we \textbf{contribute} Multi-Index Semantic Hashing (MISH), an unsupervised semantic hashing model that generates hash codes that are both effective and highly efficient through being optimized for multi-index hashing. We identify two key hash code properties for improving multi-index hashing efficiency,
related to limiting the size of the candidate set produced by multi-index hashing. We operationalize these into two novel model agnostic training objectives that effectively reduce the number of hash codes per hash table lookup, while also limiting the necessary search radius for kNN search. These new objectives are fully differentiable and enable training MISH in an end-to-end fashion, and thus enable learning hash codes highly suited for multi-index hashing. We experimentally compare MISH to state-of-the-art semantic hashing baselines in the task of document similarity search. We evaluate their efficiency by comparing the speedup obtained by multi-index hashing over a linear scan, and find that on average the baselines are upwards of 33\% slower than MISH. Even though MISH enables large efficiency gains, it is still able to obtain state-of-the-art effectiveness--summarized in Figure \ref{fig:eeplot}--where we plot the mean multi-index hashing speedup and mean prec@100 for each method relative to our MISH. Furthermore, we find that MISH can be tuned to enable even larger efficiency improvements at the cost of a slight reduction in effectiveness.  

\section{Related Work}
The problem of nearest neighbour search, also known as similarity search or proximity search, aims at finding the nearest item to a given query using a given distance measure. An efficient way of doing this is by using compact bit vectors (hash codes) that have low storage requirements and enable fast search through the use of the Hamming distance. Locality Sensitive Hashing (LSH) \cite{datar2004locality} is well-known type of hashing methods with strong theoretical guarantees \cite{wang2018survey,leskovec2020mining,datar2004locality}. However, these types of methods are \emph{data-independent}, and thus unable to capture the semantics of a document (or other types of data objects such as images). In contrast, semantic hashing \cite{salakhutdinov2009semantic} based methods are \emph{data-dependent}, and aim to learn hash codes such that nearest neighbour search on the hash codes leads to a search result similar to nearest neighbour search on the original document space \cite{wang2018survey}. 

\subsection{Semantic hashing}
Early work on semantic hashing focused on Spectral Hashing (SpH) \cite{weiss2009spectral}, an approach inspired by spectral clustering \cite{ng2002spectral} that aims at learning hash codes that preserve global similarity structures of the original documents. Laplacian co-hashing (LCH) \cite{zhang2010laplacian} learns hash codes representing document semantics through a decomposition similar to latent semantic indexing \cite{deerwester1990indexing}. Similarly to SpH, Graph Hashing \cite{liu2011hashing} uses a graph representation for learning to capture the underlying global structure. Self-Taught Hashing (STH) \cite{zhang2010self} contrasts prior work by learning to preserve the \emph{local} structures between samples identified by an initial kNN search in the original document space. The prior work above has primarily been solved as relaxed optimization problems, while later work has utilized deep learning for better capturing document semantics. Variational Deep Semantic Hashing (VDSH) \cite{chaidaroon2017variational}, the first work in this direction, uses a more complex encoding of documents through a variational autoencoder architecture (this has since become the primary architecture in subsequent work). Similarly to STH, the authors of VDSH later expanded their model (now named NbrReg) \cite{chaidaroon2018deep} by including a loss function forcing the hash codes to be able to reconstruct unique words occurring in both the document and its neighbours in the original document space (found using BM25 \cite{robertson1995okapi}). While both VDSH and NbrReg improved effectiveness, they share the problem of using a post-hoc rounding of learned real-valued vectors, rather than learning the hash codes end-to-end. To fix this, NASH \cite{nash2018} proposed to learn the hash codes end-to-end through learning to sample the bits according to a Bernoulli distribution, which reduced the quantization errors compared to a rounding approach. Based on the same principle, BMSH \cite{dong-etal-2019-document} uses a Bernoulli mixture prior, but only manages to outperform a simple version of NASH, rather than consistently outperform the full NASH model. Similarly to NbrReg and STH, recent state-of-the-art approaches have incorporated neighbourhood knowledge: RBSH \cite{hansensemhash2019} incorporates a ranking-based objective, while PairRec \cite{hansen2020PairRec} uses a pairwise reconstruction loss. The pairwise reconstruction loss is also used in our proposed MISH, and forces two hash codes of semantically similar documents to be able to reconstruct the unique words occurring in both documents, thus directly enabling the encoding of neighbourhood information into the hash codes. 

\subsection{Semantic hashing efficiency}
The semantic hashing approaches above have led to substantial improvements in effectiveness, but they all use a brute-force linear scan for doing similarity search. While this is fast due to the high efficiency of the Hamming distance, hash codes were originally developed to be used as direct indices into a hash table \cite{salakhutdinov2009semantic,norouzi2011minimal,weiss2009spectral}, as to avoid a linear scan on a dataset of potential massive size. Through a hash table, finding exact hash code matches only requires a single lookup, but when varying the search radius in a similarity search (e.g., for performing kNN search) it leads to an exponentially increasing number of lookups. To fix this, multi-index hashing \cite{greene1994multi,norouzi2012fast-initial-paper,norouzi2012fast} has been explored, which enables sub-linear search time by building hash tables on substrings of the original hash codes (see Section \ref{ss:mih} for a detailed description), and has been used for fast kNN search in hashing-based approaches related to collaborative filtering \cite{zhang2016discrete,kang2019candidate,lian2019discrete,hansen-coldstart-hash-2020,hansen2021hammingDisim}, knowledge graph search \cite{wang2019learning}, and video hashing \cite{zhang2016play}. However, none of these approaches optimize the hash codes towards improving their multi-index hashing efficiency, but rather simply apply it on already learned hash codes. In contrast, our proposed MISH is designed to directly learn hash codes suited for multi-index hashing in an end-to-end fashion, which significantly improves efficiency.

\section{Preliminaries}
\subsection{Hamming distance}
Given a document $d \in \mathcal{D}$, let $z_d \in \{-1,1\}^n$ be its associated bit string of $n$ bits, called a \emph{hash code}.
The Hamming distance between two hash codes is defined as the number of differing bits between the codes:
\begin{align}
    d_H(z_d, z_{d'}) = \sum_{i=1}^n 1_{z_{d,i} \neq z_{d',i}} = \text{SUM}(z_{d} \; \text{XOR} \; z_{d'}) 
\end{align}
where the summation can be computed efficiently due to the \textit{popcnt} instruction that counts the number of bits set to one within a machine word. Due to the efficiency of the Hamming distance, representing documents as hash codes enables both efficient radius search (retrieving all hash codes with a maximum Hamming distance of $r$ to a given hash code), as well as kNN search. Specifically, kNN is performed through radius search by incrementally increasing the search radius up until the distance to the $\text{k}^{\text{th}}$ hash code is equal to the search radius.

\subsection{Multi-index hashing}\label{ss:mih}
\citet{norouzi2012fast,norouzi2012fast-initial-paper} propose a multi-index hashing strategy for performing exact radius and kNN search in the Hamming space. The aim of multi-index hashing is to build a candidate set $\mathcal{C}$ of hash codes, significantly smaller than the full document collection, $|\mathcal{C}| \ll |\mathcal{D}|$. Given a query hash code $z$, it then suffices to compute the Hamming distances between $z$ and the hash codes within $\mathcal{C}$, rather than every hash code in the full collection $\mathcal{D}$. If the size of $\mathcal{C}$ is sufficiently small, and can be constructed efficiently, this leads to a sub-linear runtime compared to computing all possible Hamming distances.

Multi-index hashing is an efficient and easy to implement algorithm for building the candidate set $\mathcal{C}$, where each code $z \in \mathcal{D}$ is split into $m$ disjoint substrings, $z=[z^1,z^2,...,z^m]$\footnote{For ease of notation we will assume the substrings have the same length, but it is not a requirement.}. It now follows by the pigeonhole principle that if two codes $z$ and $z'$ are within radius $r$, i.e., $d_H(z,z') \leq r$, 
there exists at least one substring where the distance between the two codes is at most $r^*=\lfloor{\frac{r}{m}}\rfloor{}$. More specifically, if we assume some arbitrary fixed ordering of the substrings, and write the search radius as $r=r^*m+a, \; a<m$, one substring will have distance at most $r^*$ in the first $a+1$ substrings, or distance $r^*-1$ in the remaining $m-a-1$ substrings. Thus, the candidate set can be constructed by finding all hash codes where the distance within a substring is at most $r^*$ for the first $a+1$ substrings, or at most $r^*-1$ for the remaining $m-a-1$ substrings. For ease of notation, we will denote the substring search radius for substring $i$ as $r^*_i$.

\subsubsection{Efficient candidate set construction} \label{sss:eff-candidate-construction}
Multi-index hashing uses hash tables to construct the candidate set efficiently. It constructs $m$ hash tables, one for each substring, where the integer value of the substring is used as a key into the hash table, which then maps to all documents containing the same substring. 
Given substring $z^i$ with $\frac{n}{m}$ bits, finding exact matches would require only a single lookup, but the number of lookups for radius search scales exponentially with the substring radius $r^*_i$ as $\sum_{r'=0}^{r^*_i} (\frac{n}{m})^{r'}$. However, the exponential growth is significantly suppressed through fixing $m>1$, which reduces both the base (through the substring length) and exponent (through the substring search radius), thus making it feasible to run in practice.

Performing radius search on the hash codes can then be done in a straight-forward fashion, by searching within each substring using their associated hash table, and then taking the union over the documents found for each substring search. Note that for kNN search, incrementally increasing the search radius from $r$ to $r+1$ only changes the search radius within a single substring, and this procedure can therefore be done very efficiently by building the candidate set incrementally as $r$ is increased.



\subsubsection{Hash code properties for efficient multi-index hashing} \label{Sec:hash_code_prop}

The efficiency of multi-index hashing depends heavily on the properties of the hash codes and how they are distributed in the Hamming space. The computational cost is dominated by the cost of sorting the candidate set according to the Hamming distance to the query hash code, such that the largest speedups are obtained when the candidate set size is small. Focusing on kNN search, the candidate set size is controlled by two factors:

\begin{description}
\item[Documents per hash table lookup] 
Given a query hash code, the hash codes should be distributed such that the documents added to the candidate set are likely to appear among the top k documents with the least Hamming distance to the query hash code. To achieve this, the hash codes should be generated such that two hash codes with a low substring Hamming distance also have a low Hamming distance between the entire hash codes.

\item[Search radius for kNN] Given a query hash code, the search radius for kNN search is determined by the Hamming distance to the $\text{k}^{\text{th}}$ closest hash code, which is unknown at query time. Since the number of hash table lookups increases exponentially with the substring search radius, the hash codes should be distributed such that the Hamming distance to the $\text{k}^{\text{th}}$ document is kept low to limit the exponential growth (corresponding to substring distance less than 2).
\end{description}

Based on these factors it follows that different sets of hash codes for the same dataset can potentially have highly varying search \emph{efficiency} without necessarily affecting the search \emph{effectiveness} of the hash codes. To ensure that learned hash codes enable both efficient \emph{and} effective search, the learning procedure must reflect both of these as part of the training objective. In the next section, we present how such codes can be learned for semantic hashing.


\section{Multi-Index Semantic Hashing}
We present Multi-Index Semantic Hashing (MISH), a semantic hashing model for unsupervised semantic hashing, which learns to generate hash codes that enable \emph{both} effective and efficient search through being optimized for multi-index hashing. For a document $d \in \mathcal{D}$, MISH learns to generate an $n$-bit hash code $z_d \in \{-1,1\}^n$ that represents its semantics, such that two semantically similar documents have a low Hamming distance between them. MISH consists of a component learning to encode document semantics, and two novel components that ensure the learned hash codes are well suited for multi-index hashing (see Section \ref{ss:mih}), based on the hash code properties discussed in Section \ref{Sec:hash_code_prop}. These two components are in fact model agnostic, i.e., not tied to how the hash codes are encoded, so they are straight-forward to include in future work for improving efficiency.
Below is an overview of the components:

\begin{description}
\item[Semantic encoding] MISH is based on a variational autoencoder architecture, where the encoder learns to generate the semantic hash code according to repeating $n$ Bernoulli trials, while a decoder learns to be able to reconstruct the original document from the generated hash code.
We choose to use the pairwise reconstruction loss proposed in the state-of-the-art PairRec \cite{hansen2020PairRec} model to ensure that document semantics are well captured within the hash codes.

\item[Reducing the number of documents per hash table lookup] Given a query hash code $z_q$, during training we sample another hash code $z_s$ with a low Hamming distance within one of its substrings, but a high Hamming distance using the full hash code, which can be considered a false positive match. We derive an objective that maximises the Hamming distance between the particular substrings of $z_q$ and $z_s$, thus effectively pushing the substrings of $z_q$ and $z_s$ apart, reducing the number of such false positive matches in the hash table lookup.

\item[Control the search radius for kNN] Given a query hash code $z_q$, during training we sample a hash code $z_r$ with $d_H(z_q,z_r)=r$, where $r$ is the Hamming distance at the top k$^{\text{th}}$ position in a kNN search from $z_q$.
In case $r$ is too large, which leads to a large number of hash table lookups, $r$ is reduced through an objective that minimizes the Hamming distance between $z_q$ and $z_r$, thus effectively pushing the top k hash codes closer together.

\end{description}

\noindent In the following sections we present each component individually, and describe how they are jointly optimized for learning the hash codes in an end-to-end fashion.

\subsection{Semantic encoding} 
We use a variational autoencoder to learn a document encoder that generates a hash code capturing the document's semantics, as well as encoding the local neighbourhood structure of encoded document. This is done by training the codes such that a hash code $z$ should be able to reconstruct not only the original document $d$, but also documents in the neighborhood of $d$ defined by an appropriate similarity function. 

To learn the hash codes, we compute the log likelihood of document $d \in \mathcal{D}$ conditioned on its code $z$ as a sum of word likelihoods, which needs to be maximized:
\begin{align}
 \log p(d|z) = \sum_{j \in \mathcal{W}_d} \log p(w_j|z) 
\end{align}
where $p(z)$ is sampled by repeating $n$ Bernoulli trials and $W_d$ is the set of unique words in document $d$. However, due to the size of the Hamming space, the above is intractable to compute in practice, so the variational lower bound \cite{kingma2013auto} is maximized instead:
\begin{align}\label{trick1}
     \log p(d) \geq E_{Q(\cdot|d)}[\log p(d|z)] - \textrm{KL}( Q(z|d) || p(z))
\end{align}
where $Q(z|d)$ is a learned approximation of $p(z)$ that functions as the decoder, and $\textrm{KL}$ is the Kullback-Leibler divergence. 
In the text below, we first describe the encoder ($Q(z|d)$), then the decoder ($p(d|z)$), and lastly the loss function.

\subsubsection{Encoder}
The encoder is a feed forward network, with two hidden layers using ReLU activation units, followed by a final output layer using a sigmoid activation function, to get the bitwise sampling probabilities:
\begin{align}
Q(z|d) = \text{FF}_{\sigma}(\text{FF}_{\textrm{ReLU}}(\text{FF}_{\textrm{ReLU}}(d \odot e_{\text{imp}}))))
\end{align}
where $\text{FF}$ denotes a feed forward layer, and $e_{\text{imp}}$ is a learned word level importance embedding. The purpose of the importance embedding is to scale each word of the document representation, such that unimportant words have less influence on generating the hash codes.
During training, the bits are sampled according to the bitwise sampling probabilities, while being chosen deterministically for evaluation (choosing the most probable bit value without sampling). To make the sampling differentiable, we employ the straight-through estimator \cite{bengio2013estimating}.

\subsubsection{Decoder}
The decoder, $\log p(d|z)$, is defined as maximizing the log likelihood of each word in document $d$:
\begin{align}\label{eq:decoder_function}
    \log p(d|z) = \sum_{j \in \mathcal{W}_d} \log \frac{e^{\text{logit}(w_j|z)}}{e^{\sum_{i \in \mathcal{W}_{\textrm{all}}} \text{logit}(w_i|z)}}
\end{align}
where $\text{logit}(w|z)$ is the logit for word $w_j$ and $\mathcal{W}_{\textrm{all}}$ are all the words in the corpus. The logit for each word is computed as:
\begin{align}
    \text{logit}(w|z) = f(z)^T (E_{\textrm{word}} (I(w) \odot e_{\textrm{imp}})) + b_w
\end{align}
where $f(z)$ is a noise-infused hash code with added Gaussian noise (zero mean and a parameterized variance $\sigma^2$), which is annealed during training and results in lower variance for the gradient estimates \cite{kingma2013auto}. $E_{\textrm{word}}$ is a word embedding learned during training, $I(w)$ denotes a one-hot encoding of word $w$, and $b_w$ is a bias term.

\subsubsection{Semantic encoder loss}
To make the loss function aware of the local neighbourhood structure around a given document, we use pairwise reconstruction as proposed by \citet{hansen2020PairRec}. To this end, we use a similarity function independent of the learned hash codes to compute a set of the $p$ most semantically similar documents in the neighbourhood around document $d$, denoted as $\mathcal{N}_d^p$. For each $d_+ \in \mathcal{N}_d^p$, we construct $(d_q,d_+)$ with corresponding hash codes $(z_d,z_+)$ and define the loss function based on the variational lower bound from Eq. \ref{trick1} as:
\begin{align}\label{eq:semantic-loss}
    \mathcal{L}_{\textrm{semantic}} = -&E_{Q(\cdot |d_q)}[\log p(d_q|z_q)] + \beta\textrm{KL}( Q(z_q|d_q) || p(z_q)) \nonumber \\
    -& E_{Q(\cdot |d_{+})}[\log p(d|z_{+})] + \beta\textrm{KL}( Q(z_{+}|d_{+}) || p(z_{+}))
\end{align}
As the hash codes, $z_q$ and $z_+$ both have to reconstruct document $d_q$ (known as \emph{pairwise reconstruction}) the hash codes are forced to not only encode their associated document, but also the local neighbourhood $\mathcal{N}_d^p$ as a whole.

\subsection{Reduce the number of documents per hash table lookup}\label{ss:reduce-number-documents-lookup}



The candidate set estimated by multi-index hashing can be reduced by limiting the number of documents added by each hash table lookup. Specifically, we are interested in limiting the number of false positive matches, i.e., candidate documents added due to a low substring Hamming distance, but where the Hamming distance on the full hash code is above the search radius. Given $z_q$, a substring $i$, and the top $k$ search radii $r$ and $r^*_i$, we sample a hash code $z_s$ as follows:
\begin{align} \label{eq:reduce-size}
    z_s = \operatorname*{argmax}_{z_j} \; \; d_H(z_q,z_j) \cdot 1_{[d_H(z_q^i,z_j^i) \leq r^*_i]} \cdot 1_{[d_H(z_q,z_j) > r]} 
\end{align}
which corresponds to sampling the hash code with the largest Hamming distance that has a substring Hamming distance below $r^*_i$ and is outside the $r$-ball centered on $z_q$ (expressed via $1_{[d_H(z_q,z_j) > r]}$).
By sampling hash codes with the largest value of $d_H(z_q,z_s)$, $z_s$ is unlikely to be within top $k$, but would still appear in the candidate set due to the low substring Hamming distance. 
Based on the sampling of such hash codes, we can derive an objective that maximizes the Hamming distance within the substring as long as $z_s$ appears in the candidate set, which is expressed in the following loss function:
\begin{align}\label{eq:candiate-loss}
     \mathcal{L}_{\text{false-positive}} =  -d_H(z_q^i,z_s^i)
\end{align}
In case one or both indicator functions in Eq. \ref{eq:reduce-size} are always 0 for a given query, this loss is simply set to 0.


\subsubsection{Finding the pair $(z_q,z_s)$} \label{sss:finding-the-pair}
While training the network, the hash codes potentially change in each iteration, hence we need to continuously sample the pair $(z_q,z_s)$ during training. As recomputing every hash code for every batch is computationally expensive, we employ a memory module that is continuously updated with the generated hash codes in addition to the associated document ids. We denote this memory module as $M$. The memory size (i.e., the number of hash codes to keep in $M$) is denoted by $s_{\text{mem}}$, and the memory module is updated using a first-in first-out (FIFO) strategy. Due to the compact representation of the hash codes and document ids, we fix $s_{\text{mem}}$ to be the size of the training set\footnote{For truly massive-scale datasets, or due to specific hardware constraints, the memory size could be fixed to a number less than the training set size.}.

Using the sampling requirements from Eq. \ref{eq:reduce-size}, $z_s$ can now be obtained from the memory module. However, since the memory module may contain outdated hash codes due to model updates in previous training iterations, the validity of the pair $(z_q,z_s)$ must be ensured. This is done by recomputing $z_s$ (based on the document features obtained through the stored document id) on the current model parameters, and verifying whether it is still valid according to the sampling requirements (if not, the loss is set to 0). 
Note that the search radii $r_i^*$ and $r$ for top $k$ retrieval also needs to be estimated based on the memory module. 
However, since $z_s$ is sampled as the hash code with the largest Hamming distance to $z_q$, smaller deviations from the true radii are not problematic, as the worst-case outcome simply is that the already far apart $(z_q,z_s)$ pair is pushed slightly further apart then necessary.

\subsection{Control search radius}
In Section \ref{ss:reduce-number-documents-lookup} we detailed how to reduce the number of false positive documents per hash table lookup, while we now focus on how to reduce the number of such lookups. 
Given a query hash code $z_q$, we aim to control the search radius $r$ to limit the exponential increase in the number of hash table lookups, which happens when the substring search radius $r_i^* > 1,\; i \in \{1,...,m\}$ (see Section \ref{sss:eff-candidate-construction}), corresponding to $r > 2m-1$. To this end, we compute $r$ for the query based on the memory module, and sample a hash code $z_r$ with $d_H(z_q,z_r)=r$, resulting in the hash code pair $(z_q,z_r)$. To reduce the number of lookups, we define a loss function that minimizes the Hamming distance of the pair:
\begin{align}\label{eq:radius-loss}
      \mathcal{L}_{\text{radius}}= d_H(z_q,z_r) \cdot 1_{[r > 2m-1]}
\end{align}
where the indicator function ensures that the Hamming distance is only minimized in cases where the search radius is too large. Similarly to sampling $z_s$ for reducing the number of documents per hash table lookup (Eq. \ref{eq:reduce-size}), $z_r$ may be outdated in the memory module, but is recomputed and it is verified whether its radius is still equal to $r$ (otherwise the loss is set to 0).




\subsection{Combined loss function}
MISH is trained in an end-to-end fashion by jointly optimizing the semantic loss (Eq. \ref{eq:semantic-loss}), reducing the number of false positive documents per hash table lookup (Eq. \ref{eq:candiate-loss}), and controlling the number of such lookups (Eq. \ref{eq:radius-loss}) as follows:
\begin{align}\label{eq:total-loss}
       \mathcal{L}_{\text{total}} =  \mathcal{L}_{\textrm{semantic}} + \alpha_1 \mathcal{L}_{\text{false-positive}} + \alpha_2  \mathcal{L}_{\text{radius}}
\end{align}
where the hyperparameter weights, $\alpha_1$ and $\alpha_2$, control the trade-off between the semantic encoding and tuning the hash codes towards more efficient multi-index hashing search. However, optimizing both effectiveness and efficiency are not necessarily mutually exclusive because any permutation of the hash code bits provides the same effectiveness, but some permutations result in better multi-index hashing efficiency. Lastly, observe that $\mathcal{L}_{\text{false-positive}}$ and $\mathcal{L}_{\text{radius}}$ are model agnostic, as neither are tied to how the hash codes are generated, and can thus easily be incorporated in any semantic hashing model for improving efficiency.

\section{Experimental evaluation}
\begin{table}[]
    \centering
        \caption{Dataset statistics.}
    \resizebox{\linewidth}{!}{
    \begin{tabular}{lcccc}
        \toprule
         & documents & multi-class & classes & unique words \\ \hline
         TMC & 28,596 & Yes &  22 & 18,196 \\
         reuters & 9,848 & Yes & 90 & 16,631 \\
         agnews & 127,598 & No & 4 & 32,154 \\
        \bottomrule
    \end{tabular}
    }
    \label{tab:datasets}
\end{table}

\subsection{Datasets}
We evaluate MISH on well-known and publicly available datasets used in related work \cite{hansensemhash2019, nash2018, chaidaroon2017variational, hansen2020PairRec} and summarized in Table \ref{tab:datasets}: (1) \textit{TMC} consists of multi-class air traffic reports; (2) \textit{Agnews} consists of single-class news documents; and (3) \textit{reuters} consists of multi-class news documents, where a filtering is applied that removes documents if none of its labels occur among the top 20 most frequent labels (as done in \cite{hansensemhash2019, nash2018, chaidaroon2017variational, hansen2020PairRec}).

We use the preprocessed data provided by \citet{hansen2020PairRec}, where documents are represented using TF-IDF vectors, where words occurring only once, or in more than 90\% of the documents, are removed. We use the provided data splits, which split the datasets into training (80\%), validation (10\%), and testing (10\%), where the validation loss is used for early stopping.

\subsection{Evaluation setup}
We evaluate the hash codes in the task of document similarity search (using kNN search), where we evaluate both effectiveness and efficiency. Each document is represented as a hash code, such that document similarity search can be performed using the Hamming distance between two hash codes. For evaluation purposes, we denote a document to be relevant (i.e., similar) to a query document, if the documents share at least one label, meaning that in multi-class datasets two documents do not need to share all labels.

For effectiveness we follow related work \cite{hansensemhash2019, nash2018, chaidaroon2017variational, hansen2020PairRec} by considering the retrieval performance as measured by precision@100. 
However, existing work computes the scores based on random tie splitting, which is problematic for hash codes as ties occur often due to the limited number of $n+1$ different Hamming distances for $n$-bit hash codes. Instead, we compute the tie-aware precision@100 metric \cite{10.5555/1793274.1793325} corresponding to the average-case retrieval performance. 
Additionally, due to the large number of possible ties, we also compute the \textit{worst-case} retrieval performance by fixing ties such that irrelevant documents appear before relevant ones before computing precision@100.

For efficiency we measure the runtime for performing top-100 retrieval on the training set, where each test document acts as a query document once. We perform a linear scan, i.e., brute-force computation of all Hamming distances, as well as multi-index hashing based search \cite{norouzi2012fast}. We use the linear scan and multi-index implementation made available by \citet{norouzi2012fast}\footnote{\url{https://github.com/norouzi/mih/}}, where we follow the practical recommendation of splitting hash codes into substrings of 16 bits for multi-index hashing\footnote{\url{https://github.com/norouzi/mih/blob/master/RUN.sh}}. We repeat all timing experiments 100 times, and report the median speedup of using multi-index hashing over the linear scan. Note that the runtime of multi-index hashing naturally varies between the methods used for generating hash code, whereas the linear scan time is the same independent of the used method. All timing experiments were performed on an Intel Core i9-9940X@3.30 GHz.

\begin{table*}
    \centering
        \caption{Worst case and average case precision@100. The highest precision is highlighted in bold, and the second highest is underlined. $^{\blacktriangle}$ represents statistically significant improvements over the second best method at the 0.05 level using a two tailed paired t-test.}\vspace{-5pt}

    \resizebox{\linewidth}{!}
     {
    \begin{tabular}{l|cc|cc|cc|cc|cc|cc}
    \toprule
         &  \multicolumn{4}{c|}{Reuters} &  \multicolumn{4}{c|}{TMC} &  \multicolumn{4}{c}{Agnews} \\ 
         & \multicolumn{2}{c}{32 bits} & \multicolumn{2}{c|}{64 bits} & \multicolumn{2}{c}{32 bits} & \multicolumn{2}{c|}{64 bits} & \multicolumn{2}{c}{32 bits} & \multicolumn{2}{c}{64 bits} \\
         Prec@100 & Worst & Average & Worst & Average & Worst & Average & Worst & Average & Worst & Average & Worst & Average \\ 
         \midrule
LCH & 0.5995 & 0.6616 & 0.6283 & 0.6613 & 0.6658 & 0.7510 & 0.7421 & 0.7817 & 0.6822 & 0.7599 & 0.7423 & 0.7775\\
STH & 0.7730 & 0.8046 & 0.7803 & 0.7968 & 0.6858 & 0.7693 & 0.7481 & 0.7816 & 0.6823 & 0.8237 & 0.7931 & 0.8374\\
NbrReg & 0.5785 & 0.6329 & 0.6327 & 0.6616 & 0.3272 & 0.6648 & 0.5862 & 0.6827 & 0.7274 & 0.7914 & 0.7535 & 0.7928\\
NASH & 0.7330 & 0.7737 & 0.7767 & 0.7967 & 0.6845 & 0.7709 & 0.7535 & 0.7953 & 0.7059 & 0.8018 & 0.7748 & 0.8107\\
RBSH & 0.7809 & 0.8110 & 0.8011 & 0.8182 & \underline{0.7459} & 0.8107 & 0.7852 & 0.8158 & \underline{0.7797} & 0.8347 & 0.8053 & 0.8317\\
PairRec & \underline{0.7812} & \underline{0.8218} & \underline{0.8087} & \underline{0.8316} & 0.7320 & \textbf{0.8187} & \underline{0.7922} & \textbf{0.8288} & 0.7700 & \underline{0.8348} & \underline{0.8114} & \underline{0.8407}\\ \hline
MISH & \textbf{0.7965} & \textbf{0.8286} & \textbf{0.8248} & \textbf{0.8377} & \textbf{0.7608}$^{\blacktriangle}$ & \underline{0.8156} & \textbf{0.7931} & \underline{0.8261} & \textbf{0.7818} & \textbf{0.8375} & \textbf{0.8116} & \textbf{0.8419}\\
    \bottomrule
    \end{tabular}}
    \label{tab:resultsPerf}
\end{table*}
\begin{table*}
    \centering
        \caption{Speedup of multi-index hashing over a brute-force linear scan, as well as linear scan time per document. Greedy substring optmizing (GSO) \cite{norouzi2012fast} corresponds to the correlation-based post-hoc heuristic, and Default corresponds to using the hash codes as is. The largest speedup is highlighted in bold (independent of post-hoc fix), and the second largest is underlined. $^{\blacktriangle}$ represents statistically significant improvements  (100 repeated timing experiments) over the second best method at the 0.05 level using a two tailed paired t-test.}\vspace{-5pt}

    \resizebox{\linewidth}{!}
     {
    \begin{tabular}{l|cc|cc|cc|cc|cc|cc}
    \toprule
         &  \multicolumn{4}{c|}{Reuters} &  \multicolumn{4}{c|}{TMC} &  \multicolumn{4}{c}{Agnews} \\ 
         & \multicolumn{2}{c}{32 bits} & \multicolumn{2}{c|}{64 bits} & \multicolumn{2}{c}{32 bits} & \multicolumn{2}{c|}{64 bits} & \multicolumn{2}{c}{32 bits} & \multicolumn{2}{c}{64 bits} \\
         Speedup & Default & GSO & Default & GSO & Default & GSO & Default & GSO & Default & GSO & Default & GSO \\ \midrule
LCH & 2.5182 & 0.3508 & 0.8937 & 0.5603 & 7.2427 & 1.5879 & 2.1837 & 3.0767 & 18.2709 & 25.6365 & 3.7311 & 5.1354\\
STH & 2.9374 & 0.2783 & 1.0116 & 0.5695 & 7.0382 & 2.2211 & 2.1947 & 2.4791 & 14.6936 & 29.2779 & 4.4515 & \underline{9.5676}\\
NbrReg & \underline{6.3841} & 0.7190 & \underline{4.9589} & 0.8299 & 7.1497 & 1.7346 & 2.8800 & \underline{5.2506} & 21.7102 & 23.7698 & 6.8518 & 7.5279\\
NASH & 5.6356 & 0.6037 & 4.5869 & 0.8417 & 9.4069 & 1.5725 & 4.3819 & 4.3886 & 20.7673 & 23.1443 & 4.8355 & 5.3849\\
RBSH & 4.4342 & 0.2965 & 1.7083 & 0.8051 & 7.0831 & 1.9708 & 2.7957 & 2.9146 & 25.9186 & 27.1164 & 4.5036 & 4.7345\\
PairRec & 5.4296 & 0.3721 & 3.0770 & 0.9433 & \underline{10.9118} & 1.7063 & 5.1129 & 4.9614 & 29.7880 & \underline{33.4096} & 7.1765 & 7.8676\\ \hline
MISH & \textbf{7.0698}$^{\blacktriangle}$& 1.1216 & \textbf{5.4466}$^{\blacktriangle}$& 1.0282 & \textbf{14.6296}$^{\blacktriangle}$& 2.1923 & \textbf{8.8696}$^{\blacktriangle}$& 6.1645 & \textbf{44.0151}$^{\blacktriangle}$& 35.9177 & \textbf{13.6756}$^{\blacktriangle}$& 12.5788\\
    \bottomrule
Linear scan time & \multicolumn{2}{c|}{0.000070 s} & \multicolumn{2}{c|}{0.000069 s} & \multicolumn{2}{c|}{0.000111 s} & \multicolumn{2}{c|}{0.000109 s} & \multicolumn{2}{c|}{0.000432 s} & \multicolumn{2}{c|}{0.000407 s} \\ \bottomrule
    \end{tabular}}
    \label{tab:resultsSpeed}
\end{table*}

\subsection{Baselines}
We compare our proposed MISH to non-neural approaches with post-processing rounding of document vectors to obtain hash codes (STH \cite{zhang2010self} and LCH \cite{zhang2010laplacian}), neural approaches with post-processing rounding (NbrReg \cite{chaidaroon2018deep}), and neural end-to-end approaches that incorporate the rounding as part of the model (NASH \cite{nash2018}, RBSH \cite{hansensemhash2019}, and PairRec \cite{hansen2020PairRec}). All baselines are tuned following the original papers.

In contrast to our MISH, existing semantic hashing baselines have focused on maximizing effectiveness, while simply assumed retrieval is done using a brute-force linear scan, rather than faster alternatives such as multi-index hashing.
However, how bits are assigned into substrings impacts multi-index hashing efficiency, as the candidate set size may be larger than necessary. To this end, we include the greedy substring optimization (GSO) heuristic proposed by \citet{norouzi2012fast}, which greedily assigns bits to substrings as to minimize the correlation between the bits within each substring.

\subsection{Tuning}
To tune MISH\footnote{We make our code publicly available at \url{https://github.com/Varyn/MISH}} we fix the number of hidden units in the encoder to 1000, and vary the number of documents in the pairwise reconstruction neighbourhood ($\mathcal{N}_d^p$) from $\{10,25,50,100\}$, where both 10 and 25 worked well for reuters and TMC, whereas 100 was consistently chosen for agnews. Similarly to \citet{hansen2020PairRec}, $\mathcal{N}_d^p$ is constructed based on retrieving the top $p$ most semantically similar documents based on 64 bit STH hash codes. For the KL-divergence, we tune $\beta$ from $\{0, 0.01\}$, where 0 was chosen most often, thus effectively removing the KL term in those cases. For the variance annealing in the noise-infused hash codes, we fix the initial value to 1 and reduce by $10^{-6}$ every iteration (as per \cite{hansen2020PairRec,hansensemhash2019}). For the combined loss, we tune $\alpha_1$ from $\{1,3,5,7\}$ and $\alpha_2$ from $\{0.01,0.05,0.1\}$, where $\alpha_1=3$ and $\alpha_2=0.01$ was chosen for reuters and TMC, and $\alpha_1=7$ and $\alpha_2=0.05$ for agnews. As we focus on learning hash codes that maintain state-of-the-art effectiveness, while improving efficiency, we choose to use only the semantic loss ($\mathcal{L}_{\textrm{semantic}}$) on the validation set for model selection and early stopping, rather than the weighted total loss.
Lastly, MISH is optimized using the Adam optimizer \cite{kingma2014adam} with a learning rate from $\{0.001, 0.005\}$, where 0.005 was chosen for reuters and TMC, and 0.001 for agnews. 

\subsection{Results}
The experimental results are summarized for effectiveness in Table \ref{tab:resultsPerf} and efficiency in Table \ref{tab:resultsSpeed}. In Table \ref{tab:resultsPerf}, the highest worst-case and average-case scores per column are highlighted in bold, and the second highest are underlined. In Table \ref{tab:resultsSpeed}, the largest and second largest speedups (independent of applying the greedy substring optimization (GSO)) are highlighted in bold and underlined, respectively. Additionally, we report the linear scan time per document as a point of reference for the speedups. 
In both tables, $^{\blacktriangle}$ represents statistically significant improvements over the second best method at the 0.05 level using a two -tailed paired t-test.

\subsubsection{Retrieval effectiveness}
Table \ref{tab:resultsPerf} shows the effectiveness measured by worst-case and average-case prec@100 across the datasets using 32 and 64 bit hash codes (corresponding to the typical machine word sizes). Across all methods, we observe a larger gain in worst-case prec@100 when increasing the number of bits in the hash codes, compared to the average-case prec@100, where only smaller increases are obtained. Thus, increasing the number of bits is beneficial when worst-case performance is important no matter the chosen method. The increase in worst-case prec@100 happens because the documents are being spread out more in the Hamming space as the number of bits are increased, which reduce the number of Hamming distance ties.

Our MISH method obtains the best results for worst-case prec@100 in all cases, and additionally it obtains the best average-case prec@100 for reuters and agnews. For TMC, PairRec obtains marginally higher average-case prec@100 compared to MISH, but PairRec is in most cases the second best method. In general, the difference in effectiveness between the best and second best performing method is relatively small, and we only obtain statistically significant improvements for worst-case prec@100 at 32 bits for TMC. As both MISH and the state-of-the-art PairRec are based on the same semantic loss (Eq. \ref{eq:semantic-loss}), it was to be expected that MISH would not significantly improve effectiveness over PairRec.
However, it is important to notice that including the two additional losses in MISH, for improving efficiency, did not negatively impact effectiveness either. In fact, the additional losses have a regularizing effect that reduces the number of Hamming distance ties (hence improving the worst-case performance), as the losses force the hash codes to be better spread in the Hamming space.

Table \ref{tab:deltagap} shows the average percentage decreases in prec@100 (for both worst-case and average-case) compared to the best worst-case and average-case scores, respectively. For a given method, a decrease of $0\%$ corresponds to that method always being the best performing method across all datasets. We observe that on average, MISH outperforms the other methods in both cases, with a noticeable better average worst-case effectiveness. Additionally, it can be seen that the baselines generally have larger worst-case decreases compared to the average-case decreases, which shows that the baselines broadly cluster the hash codes more, thus resulting in larger number of Hamming distance ties.

\begin{table}
    \centering
        \caption{Average decrease in worst-case and average-case prec@100 compared to the best  scores per dataset. An average decrease of 0\% corresponds to obtaining the best performance across all datasets.\label{tab:deltagap}}
        \vspace{-5pt}
        \begin{tabular}{l|cc|cc}
    \toprule
    & \multicolumn{2}{c}{32 bits} & \multicolumn{2}{c}{64 bits} \\ 
    & $\Delta$ Worst & $\Delta$ Average & $\Delta$ Worst & $\Delta$ Average \\ \midrule
LCH & -16.65\% & -12.57\% & -12.93\% & -11.46\% \\
STH & -8.51\% & -3.53\% & -4.45\% & -3.71\% \\
NbrReg & -30.44\% & -15.98\% & -18.84\% & -14.83\% \\
NASH & -9.24\% & -5.58\% & -5.12\% & -4.21\% \\
RBSH & -1.39\% & -1.15\% & -1.55\% & -1.71\% \\
PairRec & -2.40\% & -0.38\% & -0.70\% & -0.29\% \\ \hline
MISH & \textbf{0.00\%} & \textbf{-0.13\%} & \textbf{0.00\%} & \textbf{-0.11\%} \\
    \bottomrule
    \end{tabular}
    \vspace{-11pt}
  \end{table}
  
  \begin{figure}
    \includegraphics[width=1\linewidth]{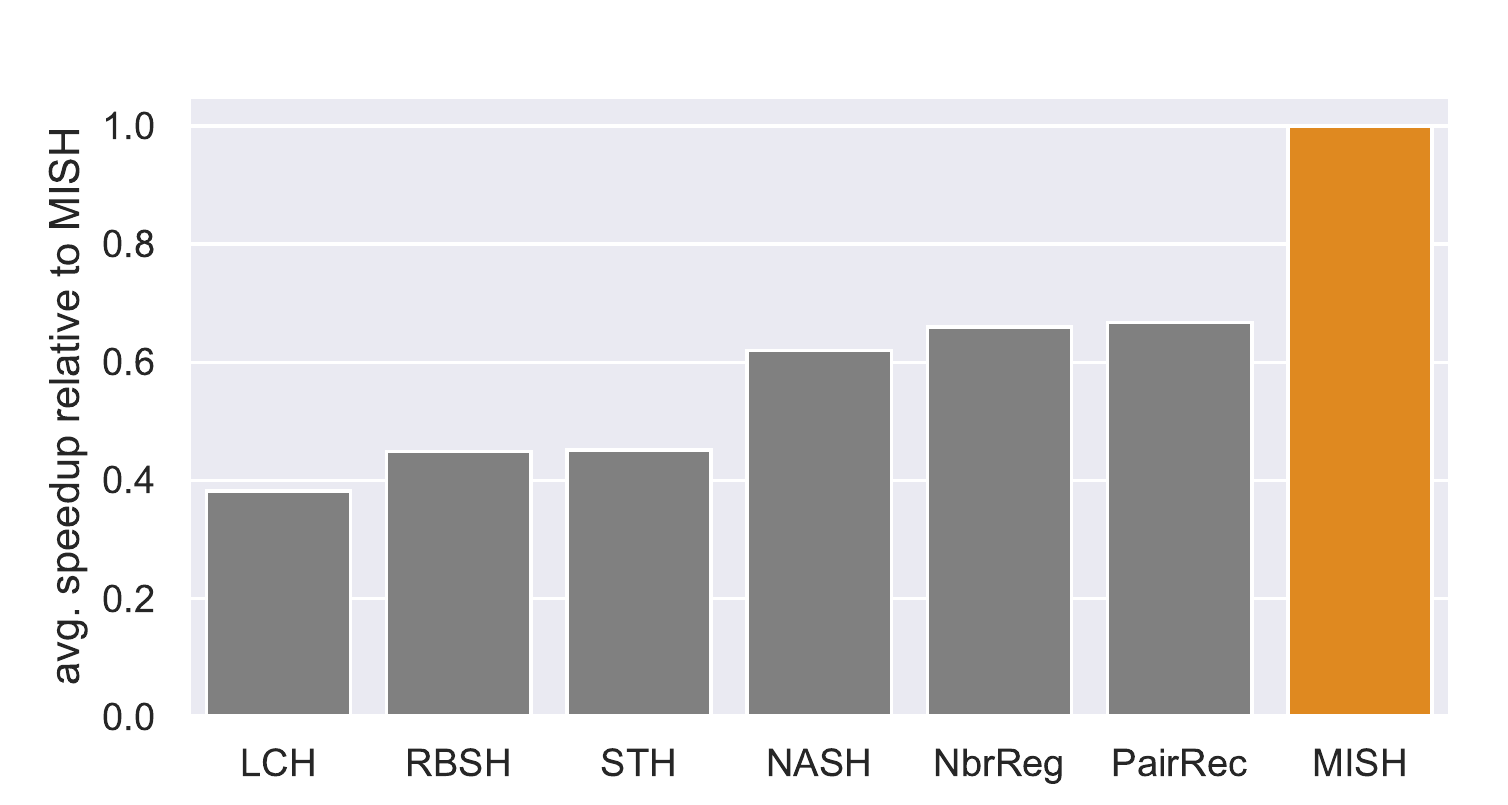}
    \vspace{-20pt}
    \caption{Speedup of multi-index hashing over linear scan relative to MISH averaged across all datasets and bit configurations.\label{fig:speedup}}
  \vspace{-5pt}
\end{figure}

\subsubsection{Retrieval efficiency}
Table \ref{tab:resultsSpeed} shows the relative speedup compared to a linear scan of the hash codes produced by each method, with and without greedy substring optimization (GSO) \cite{norouzi2012fast}, together with the linear scan time per document. The linear scan time per document is slightly lower for 64 bit hash codes compared to 32 bit, due to our machine using a 64 bit operating system. In addition, due to the sequential access pattern in a linear scan, the larger memory requirement of 64 bit hash codes does not increase the scan time. Overall, we observe that all methods achieve a higher speedup for 32 bits compared to the speedup at 64 bits, caused by an increase in the search radius for top 100 retrieval, which leads to a larger number of hash table lookups, thus increasing the candidate set of multi-index hashing. Furthermore, as expected the speedup increases as the dataset size increases, because the relative size of the candidate set decreases compared to the entire set of hash codes.

Table \ref{tab:resultsSpeed} shows that MISH is significantly faster than the baselines on all datasets and number of bits. The speedup obtained by the baselines are largely varied across the datasets and number of bits. No baseline can perform consistently well in all settings. GSO leads to larger speedups for the baselines on agnews and for 64 bit hash codes on TMC (except for PairRec), but worse on reuters and 32 bit hash codes on TMC. For MISH, applying GSO always decreases the speedup, which is caused by GSO changing the substring structure learned during training through our proposed loss functions (Eq. \ref{eq:candiate-loss} and Eq. \ref{eq:radius-loss}), to a less optimal one.
This highlights that GSO is not always beneficial and, more importantly, the limitation of making a post-hoc change to the structure of the hash codes. In contrast, MISH directly optimizes the desirable hash code properties for multi-index hashing (see Section \ref{Sec:hash_code_prop}) in an end-to-end fashion, which significantly improves multi-index hashing efficiency.

Figure \ref{fig:speedup} shows the average relative speedup, across all datasets and number of bits, compared to MISH. Generally the baselines can be clustered in two groups of similar average efficiency (LCH, RBSH, STH) and (NASH, NbrReg, PairRec). Interestingly, RBSH is the only neural method among the least efficient methods, even though PairRec and RBSH share the same underlying neural architecture with the difference being that RBSH uses a pairwise ranking loss and PairRec uses a pairwise reconstruction loss. This shows that it is problematic to assume the produced hash codes by a given method will be efficient, without directly optimizing them as done in MISH, as even small changes in the model may greatly influence the efficiency.

\begin{figure}
    \centering
    \includegraphics[width=0.49\linewidth]{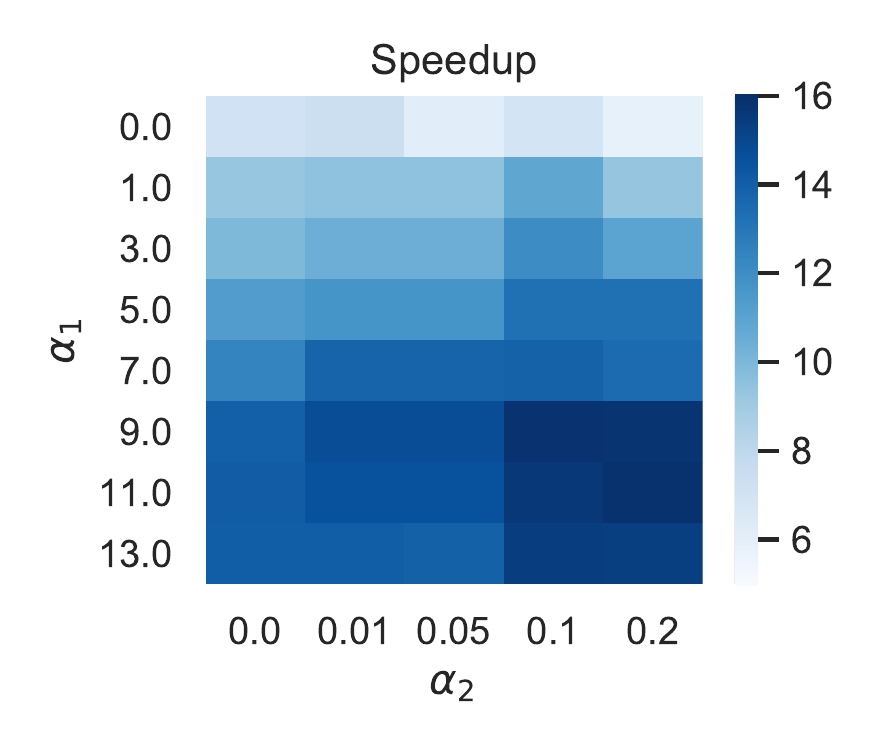}
    \includegraphics[width=0.49\linewidth]{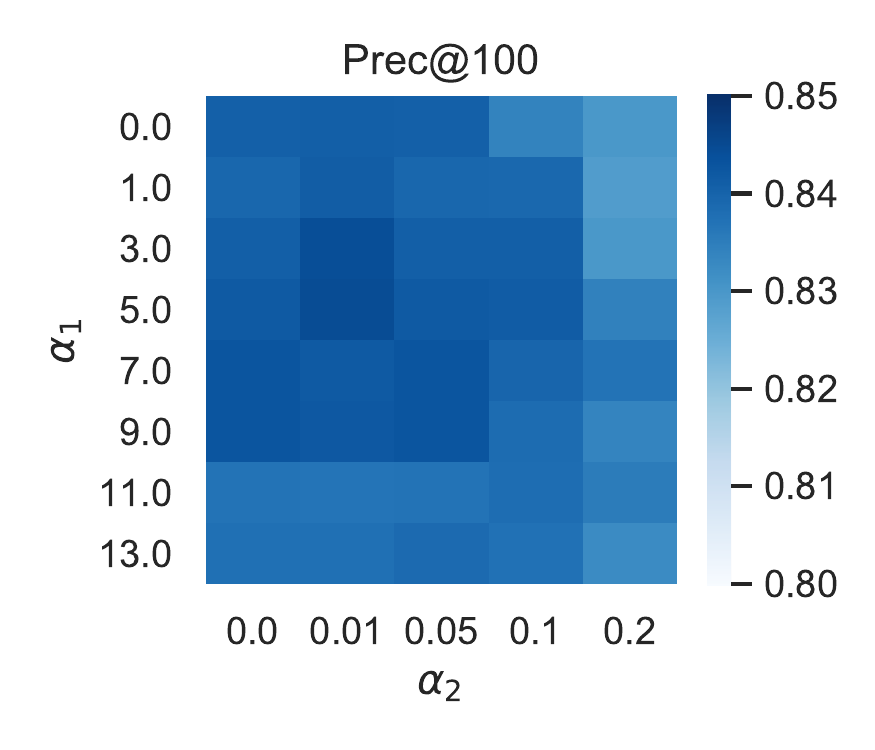}
    \vspace{-15pt}
    \caption{Hyperparameter impact on speedup and prec@100: $\alpha_1$ reduces the number of documents per hash table lookup, and $\alpha_2$ controls the number of such lookups.}
    \label{fig:paramgrid}
\end{figure}

\subsection{Efficiency and effectiveness impact of $\alpha_1$ and $\alpha_2$}
To maintain state-of-the-art effectiveness, model selection was done based only on the semantic loss (Eq. \ref{eq:semantic-loss}) on the validation set, but model training is naturally still done on the weighted total loss (Eq. \ref{eq:total-loss}).
We now investigate the impact of the total loss weights ($\alpha_1$ and $\alpha_2$) on the efficiency and effectiveness of MISH, where $\alpha_1$ reduces the number of documents per hash table lookup, and $\alpha_2$ controls the number of such lookups. To this end, we report the speedup and average-case prec@100 for as two grid plots with $\alpha_1$ and $\alpha_2$ as the axes, exemplified for 64 bit hash codes on agnews.

The grid plots can be seen in Figure \ref{fig:paramgrid}, where the top left corner ($\alpha_1=0, \; \alpha_2=0$) corresponds to the PairRec baseline \cite{hansen2020PairRec}. For the speedup plot, we observe a clear trend that higher values of both $\alpha_1$ and $\alpha_2$ improve efficiency, but $\alpha_1$ has the largest impact. This is expected since $\alpha_1$ directly reduces the number of documents per hash table lookup,  reducing the candidate set across all queries, while $\alpha_2$ affects a smaller subset of queries who exhibit a large search radius. For the prec@100 plot, we observe that higher values of $\alpha_1$ and $\alpha_2$ reduce effectiveness, which highlights the possible trade-off when tuning the hyperparameters. However, the area with the largest prec@100 scores is relatively large, thus enabling a large efficiency improvement without compromising effectiveness.



\begin{figure*}
    \centering
    \includegraphics[width=0.33\linewidth]{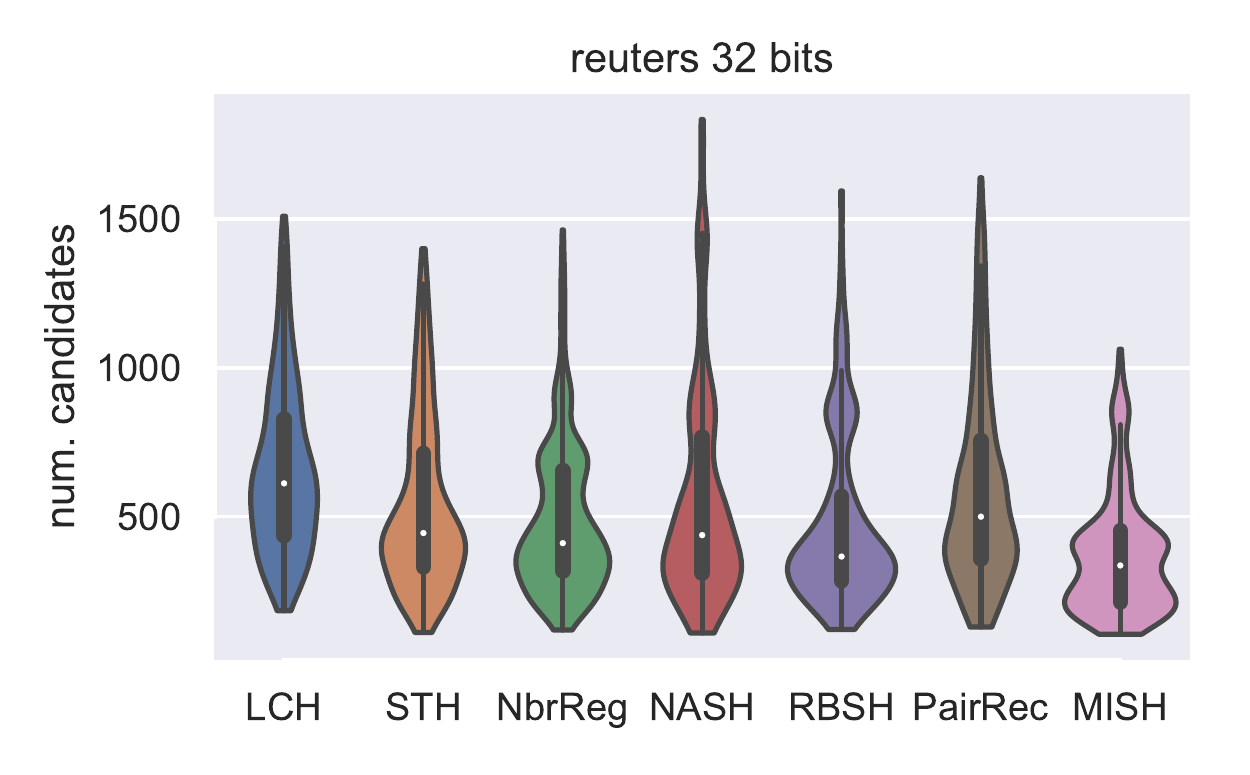}
    \includegraphics[width=0.33\linewidth]{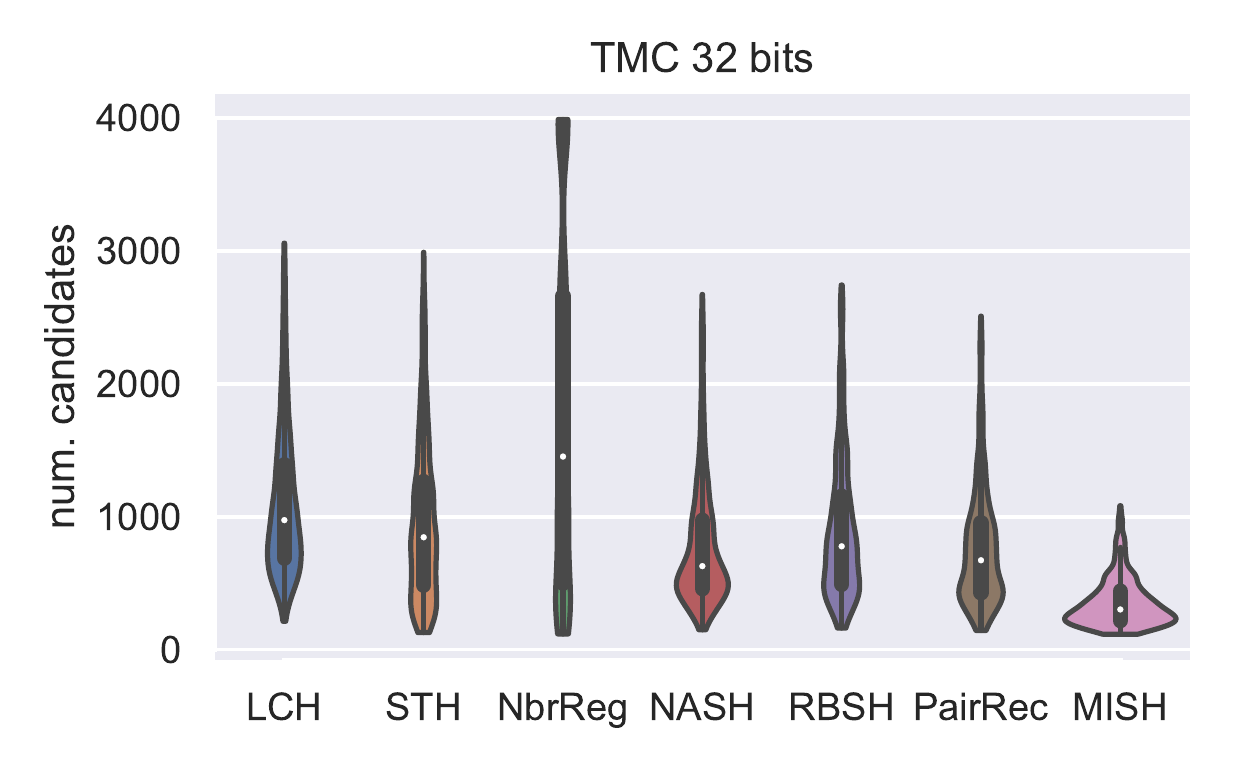}
    \includegraphics[width=0.33\linewidth]{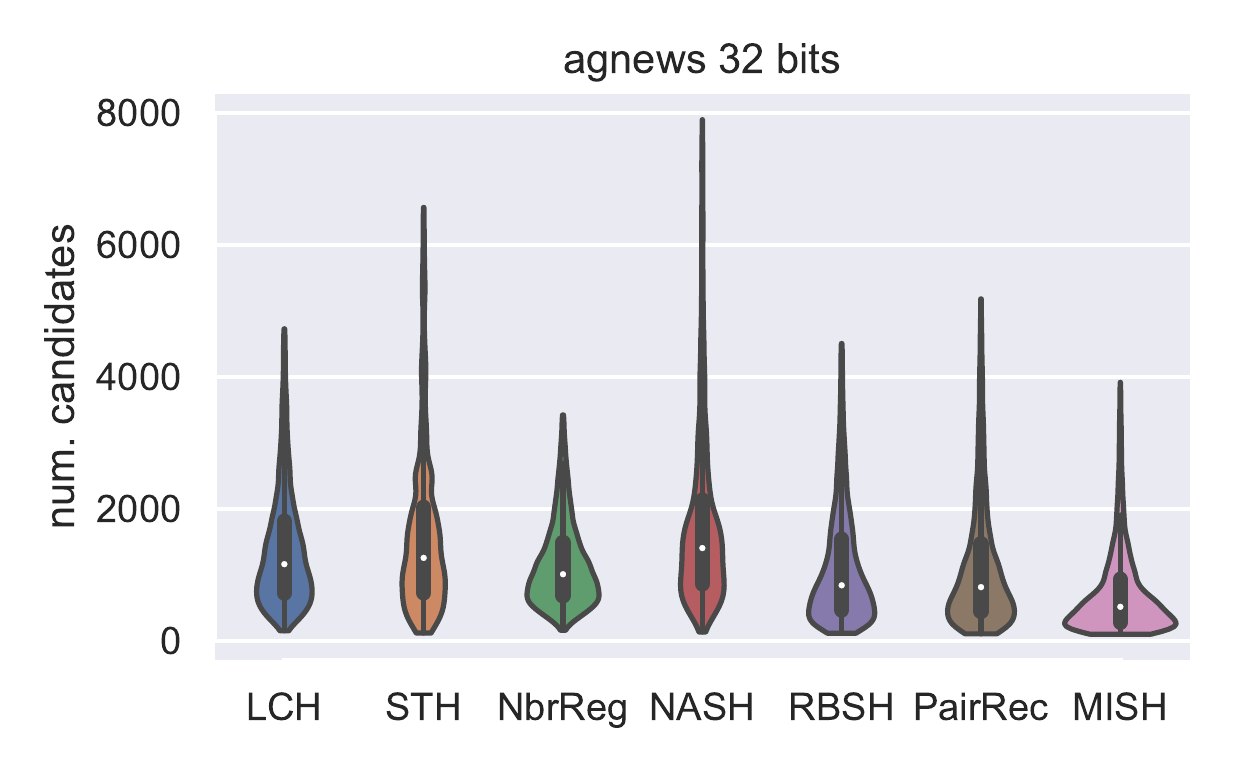}
    \includegraphics[width=0.33\linewidth]{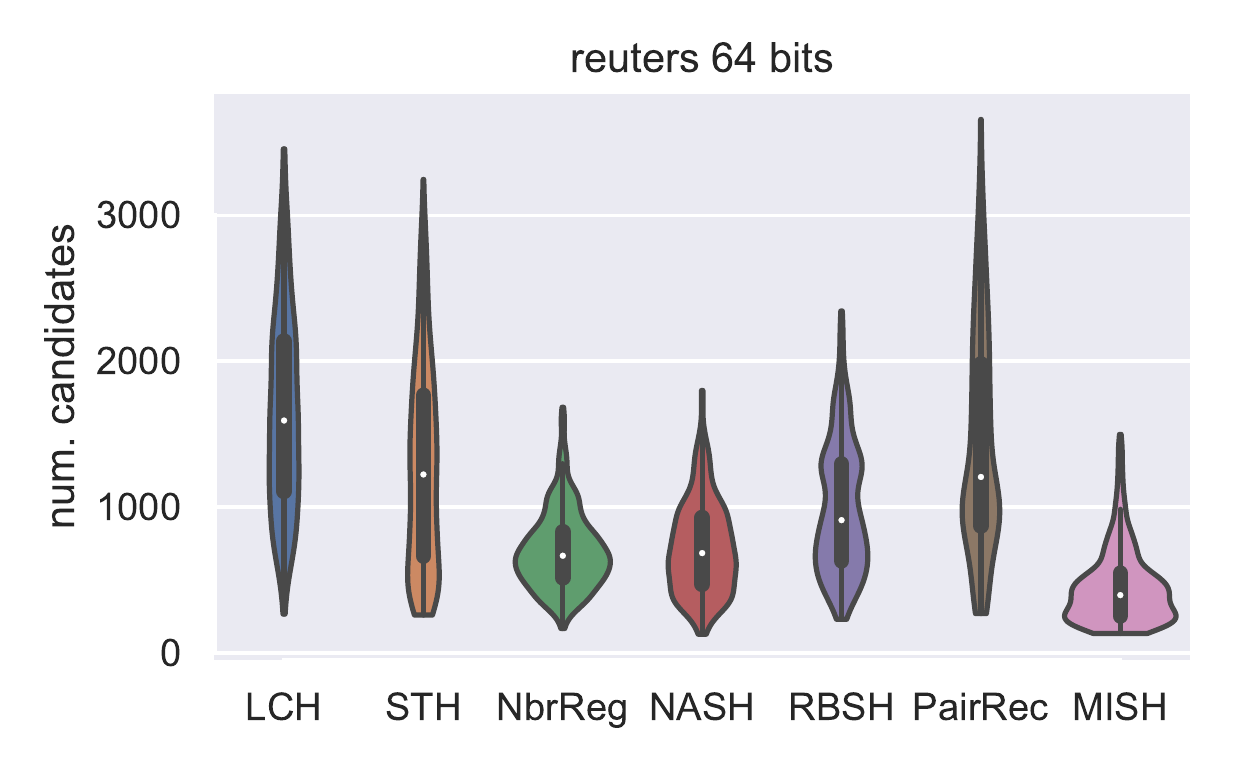}
    \includegraphics[width=0.33\linewidth]{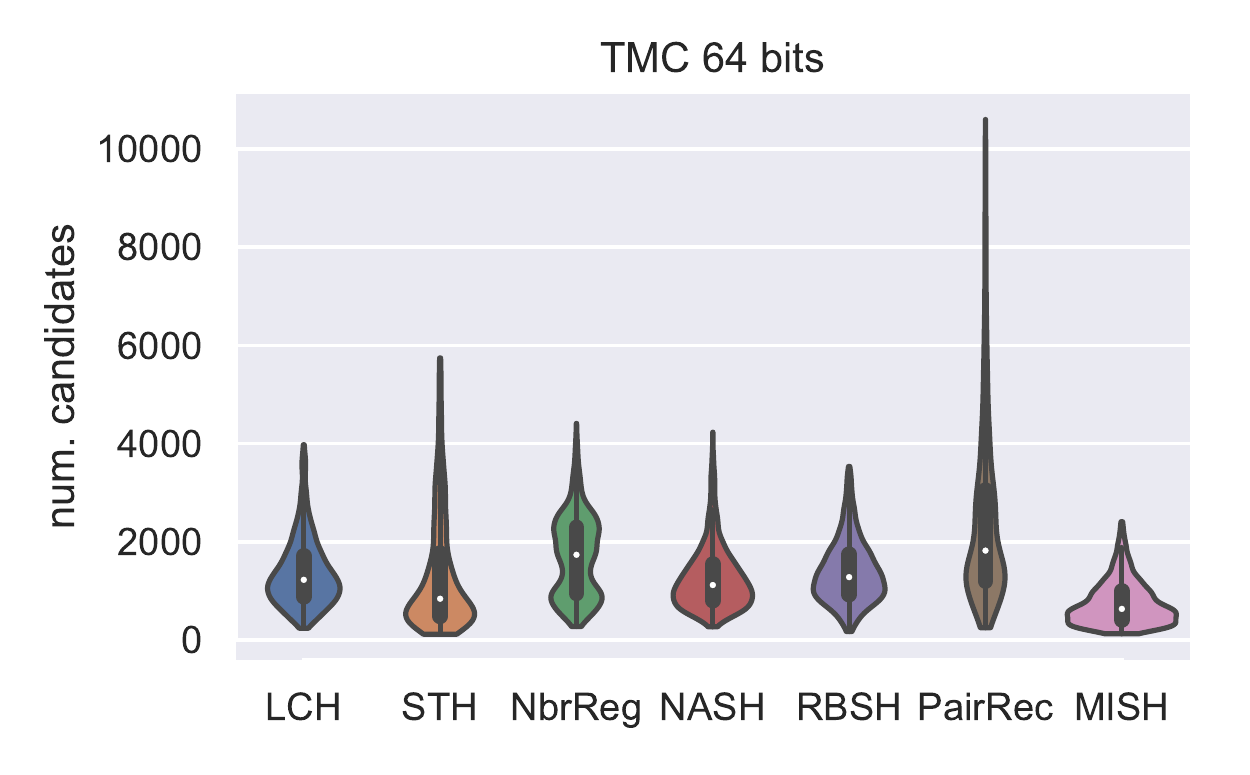}
    \includegraphics[width=0.33\linewidth]{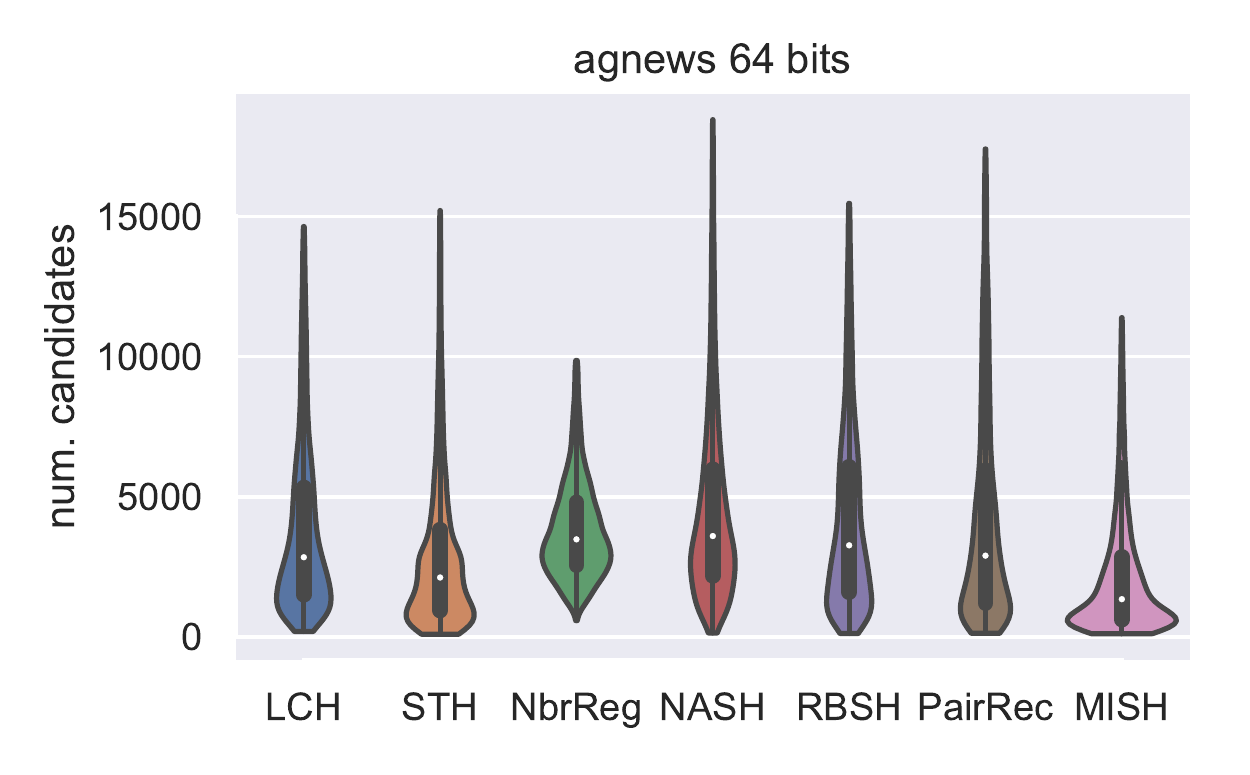}
    \caption{Violin plots (combined density and boxplot) of the number of document candidates found using multi-index hashing for top-100 retrieval.}
    \label{fig:candiates}
\end{figure*}

\begin{figure*}
    \centering
\includegraphics[width=0.19\linewidth]{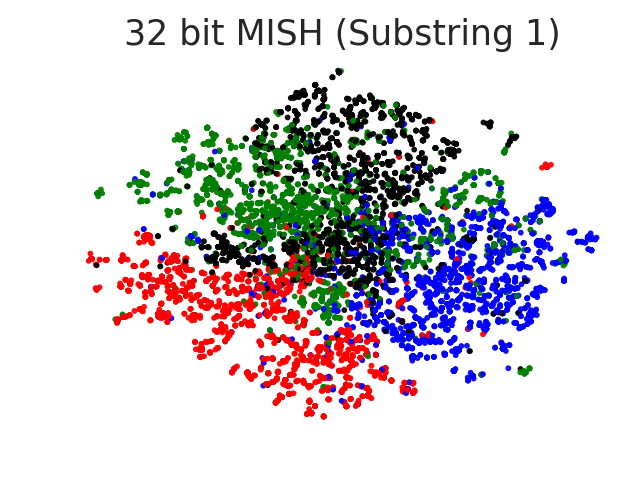}
\includegraphics[width=0.19\linewidth]{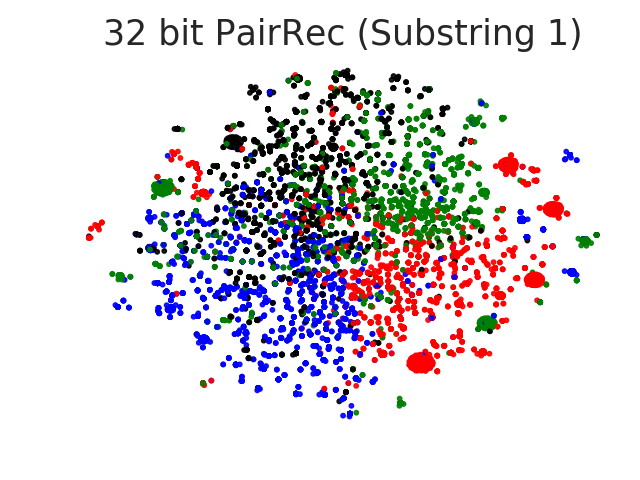}
\includegraphics[width=0.19\linewidth]{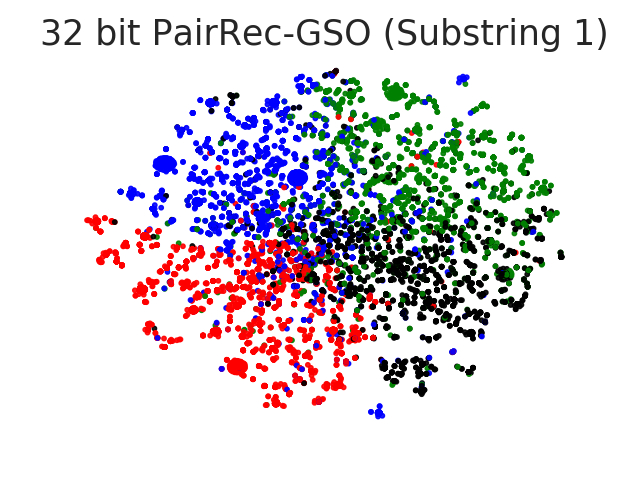}
\includegraphics[width=0.19\linewidth]{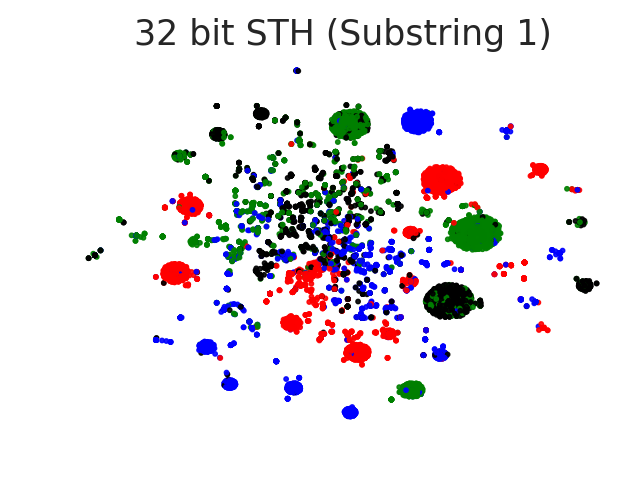}
\includegraphics[width=0.19\linewidth]{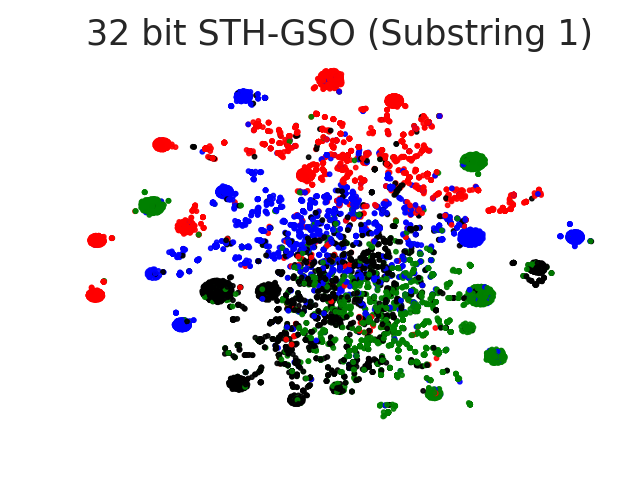}

\includegraphics[width=0.19\linewidth]{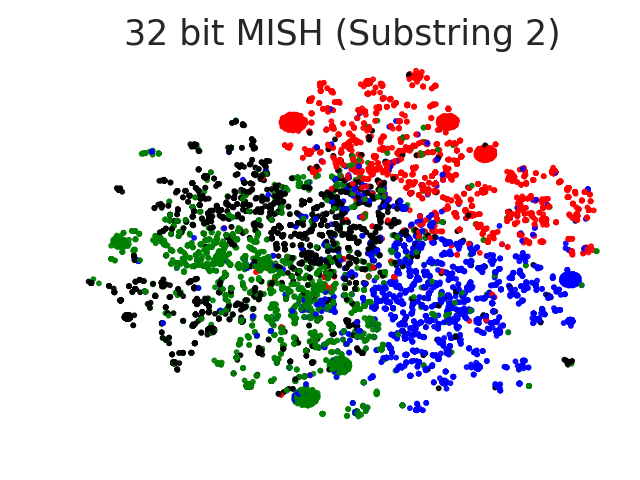}
\includegraphics[width=0.19\linewidth]{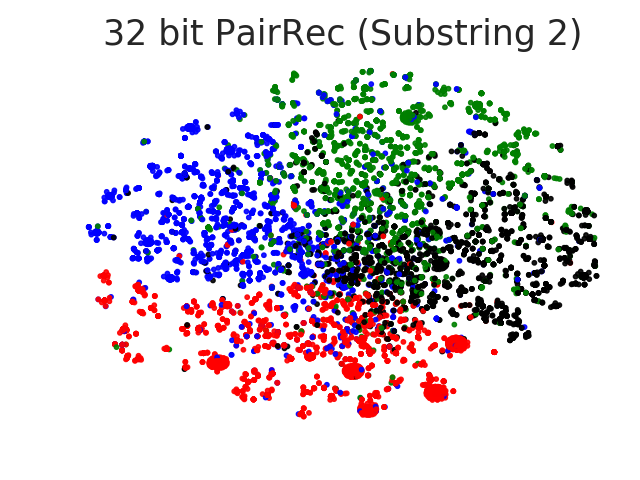}
\includegraphics[width=0.19\linewidth]{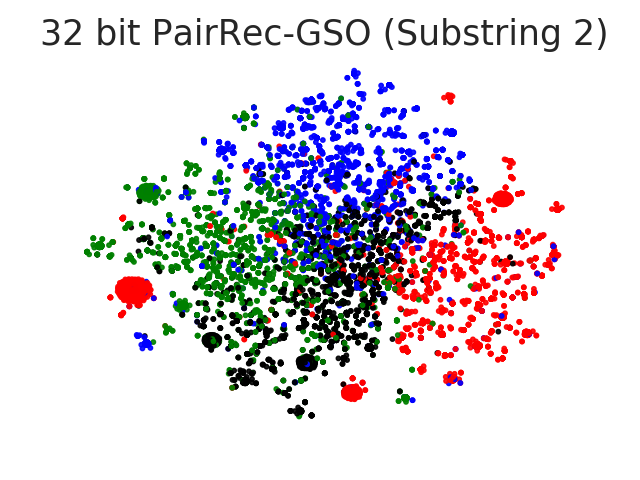}
\includegraphics[width=0.19\linewidth]{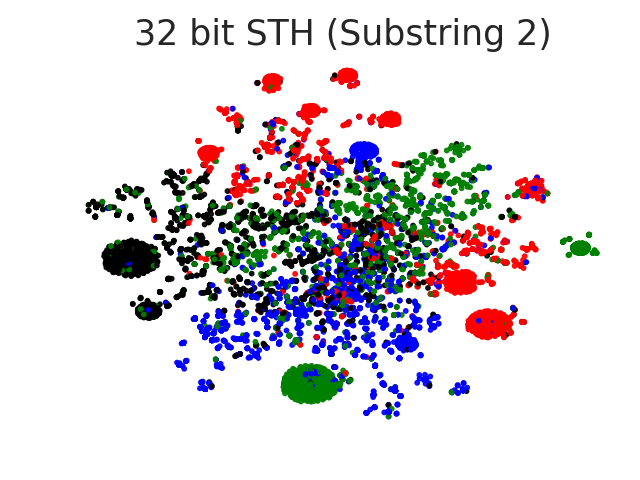}
\includegraphics[width=0.19\linewidth]{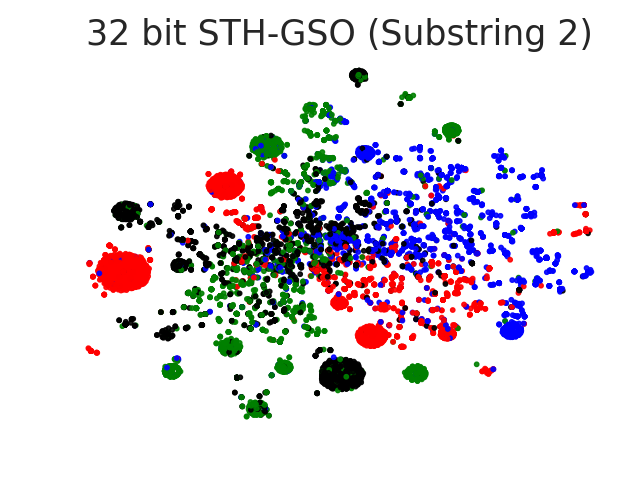}
    \caption{t-SNE \cite{maaten2008visualizing} visualization of the two substring of MISH, PairRec, and STH 32 bit hash codes on agnews. Each color represents one of 4 different classes. Greedy substring optimization (GSO) \cite{norouzi2012fast} is applied on PairRec and STH as it improved their effiency for multi-index hashing.}
    \label{fig:visualition}
\end{figure*}
\subsection{Distribution of candidate set sizes}

The computational cost of multi-index hashing is dominated by the cost of sorting the candidate set according to the Hamming distances to the query hash code. We now investigate the distribution of the candidate set size per hash code query for each method, as visualized in Figure \ref{fig:candiates} using symmetrical violin plots\footnote{A violin plot is a combination of a boxplot and a symmetrical density plot that shows the full distribution of the data.}. For the cases where GSO leads to improved speedups, we compute the candidate set based on the hash codes after applying GSO. Across all datasets and bit sizes, we observe that MISH is able to better concentrate the density towards a lower number of candidates compared to the baselines, thus explaining the large speedup improvements. By observing the larger candidate sizes for 64 bit hash codes compared to 32 bit hash code, we can directly see the reason for the speedup gap between the two hash code sizes reported in Table \ref{tab:resultsSpeed}. Furthermore, the baselines exhibiting poor speedups are primarily due to long density tails or a more uniform distribution of candidate set sizes. While this naturally leads to worse overall efficiency, it also has the potential problem of large query time variance, which may limit their application in extremely time-constrained use cases.

\subsection{Substring-level visualization}
To further investigate the differences in how the methods distribute the bits within the hash codes, we choose to visualize the individual substrings. We consider 32 bit hash codes from the test set of agnews, as it only contains two substrings, while agnews is a single-class dataset, which makes it easier to visualize if hash codes of the same class are clustered. We consider MISH, PairRec, and STH as representative methods, as it also enables visualizing the impact of GSO on the two baselines. PairRec is chosen due to being the second best method on both effectiveness and efficiency for 32 bit hash codes on agnews, while STH represents a method where GSO greatly improves (doubles) its speedup (see Table \ref{tab:resultsSpeed}). Figure \ref{fig:visualition} shows a two-dimensional t-SNE \cite{maaten2008visualizing} visualization, where it is important to keep in mind that each plot contains the same number of points, such that a plot appearing more sparse (more distance between points) corresponds to more hash codes being highly similar. When comparing MISH and PairRec, we observe that they do appear similar, but MISH is slightly less sparse, meaning the hash codes are better spread throughout the Hamming space. For STH, we observe that prior to applying GSO, the hash codes are tightly clustered and highly sparse (especially in the first substring), but GSO is able to redistribute the bits such that the substrings better utilize the space. This redistribution reduces the amount of false positive candidates found in multi-index hashing, thus leading to the large observed speedup.


\subsection{Conclusion}
We presented Multi-Index Semantic Hashing (MISH), an unsupervised semantic hashing model that learns hash codes well suited for multi-index hashing \cite{norouzi2012fast}, which enables highly efficient document similarity search. Compared to a brute-force linear scan over all the hash codes, multi-index hashing constructs a smaller candidate set to search over, which can provide sub-linear search time. We identify key hash code properties that affect the size of the candidate set, and use them to derive two novel objectives that enable MISH to learn hash codes that results in smaller candidate sets when using multi-index hashing. Our objectives are model agnostic, i.e., not tied to how the hash codes are generated specifically in MISH, which means they are straight-forward to incorporate in existing and future semantic hashing models.
We experimentally compared MISH to state-of-the-art semantic hashing baselines in the task of document similarity search, where we evaluated both efficiency and effectiveness. While multi-index hashing also improves the efficiency of the baseline hash codes compared to a linear scan, they are still upwards of 33\% slower than our proposed MISH. Interestingly, these large efficiency gains of MISH can be obtained without reducing effectiveness, as MISH is still able to obtain state-of-the-art effectiveness, but we do find that even further efficiency improvements can be obtained, but at the cost of an effectiveness reduction.
In future work, we plan to explore supervised versions of MISH, specifically the impact of expanding our proposed efficiency objectives with label information, which could decrease the number of irrelevant documents in the candidate sets.

\balance
\bibliographystyle{ACM-Reference-Format}
\bibliography{bibfile.bib}


\begin{thebibliography}{35}


\ifx \showCODEN    \undefined \def \showCODEN     #1{\unskip}     \fi
\ifx \showDOI      \undefined \def \showDOI       #1{#1}\fi
\ifx \showISBNx    \undefined \def \showISBNx     #1{\unskip}     \fi
\ifx \showISBNxiii \undefined \def \showISBNxiii  #1{\unskip}     \fi
\ifx \showISSN     \undefined \def \showISSN      #1{\unskip}     \fi
\ifx \showLCCN     \undefined \def \showLCCN      #1{\unskip}     \fi
\ifx \shownote     \undefined \def \shownote      #1{#1}          \fi
\ifx \showarticletitle \undefined \def \showarticletitle #1{#1}   \fi
\ifx \showURL      \undefined \def \showURL       {\relax}        \fi
\providecommand\bibfield[2]{#2}
\providecommand\bibinfo[2]{#2}
\providecommand\natexlab[1]{#1}
\providecommand\showeprint[2][]{arXiv:#2}

\bibitem[\protect\citeauthoryear{Bengio, L{\'e}onard, and Courville}{Bengio
  et~al\mbox{.}}{2013}]%
        {bengio2013estimating}
\bibfield{author}{\bibinfo{person}{Yoshua Bengio}, \bibinfo{person}{Nicholas
  L{\'e}onard}, {and} \bibinfo{person}{Aaron Courville}.}
  \bibinfo{year}{2013}\natexlab{}.
\newblock \showarticletitle{Estimating or propagating gradients through
  stochastic neurons for conditional computation}.
\newblock \bibinfo{journal}{\emph{arXiv preprint arXiv:1308.3432}}
  (\bibinfo{year}{2013}).
\newblock


\bibitem[\protect\citeauthoryear{Chaidaroon, Ebesu, and Fang}{Chaidaroon
  et~al\mbox{.}}{2018}]%
        {chaidaroon2018deep}
\bibfield{author}{\bibinfo{person}{Suthee Chaidaroon}, \bibinfo{person}{Travis
  Ebesu}, {and} \bibinfo{person}{Yi Fang}.} \bibinfo{year}{2018}\natexlab{}.
\newblock \showarticletitle{Deep Semantic Text Hashing with Weak Supervision}.
  SIGIR, \bibinfo{pages}{1109--1112}.
\newblock


\bibitem[\protect\citeauthoryear{Chaidaroon and Fang}{Chaidaroon and
  Fang}{2017}]%
        {chaidaroon2017variational}
\bibfield{author}{\bibinfo{person}{Suthee Chaidaroon} {and} \bibinfo{person}{Yi
  Fang}.} \bibinfo{year}{2017}\natexlab{}.
\newblock \showarticletitle{Variational deep semantic hashing for text
  documents}. In \bibinfo{booktitle}{\emph{SIGIR}}. \bibinfo{pages}{75--84}.
\newblock


\bibitem[\protect\citeauthoryear{Datar, Immorlica, Indyk, and Mirrokni}{Datar
  et~al\mbox{.}}{2004}]%
        {datar2004locality}
\bibfield{author}{\bibinfo{person}{Mayur Datar}, \bibinfo{person}{Nicole
  Immorlica}, \bibinfo{person}{Piotr Indyk}, {and} \bibinfo{person}{Vahab~S
  Mirrokni}.} \bibinfo{year}{2004}\natexlab{}.
\newblock \showarticletitle{Locality-sensitive hashing scheme based on p-stable
  distributions}. In \bibinfo{booktitle}{\emph{Proceedings of the twentieth
  annual symposium on Computational geometry}}. ACM, \bibinfo{pages}{253--262}.
\newblock


\bibitem[\protect\citeauthoryear{Deerwester, Dumais, Furnas, Landauer, and
  Harshman}{Deerwester et~al\mbox{.}}{1990}]%
        {deerwester1990indexing}
\bibfield{author}{\bibinfo{person}{Scott Deerwester}, \bibinfo{person}{Susan~T
  Dumais}, \bibinfo{person}{George~W Furnas}, \bibinfo{person}{Thomas~K
  Landauer}, {and} \bibinfo{person}{Richard Harshman}.}
  \bibinfo{year}{1990}\natexlab{}.
\newblock \showarticletitle{Indexing by latent semantic analysis}.
\newblock \bibinfo{journal}{\emph{Journal of the American society for
  information science}} \bibinfo{volume}{41}, \bibinfo{number}{6}
  (\bibinfo{year}{1990}), \bibinfo{pages}{391--407}.
\newblock


\bibitem[\protect\citeauthoryear{Dong, Su, Shen, and Chen}{Dong
  et~al\mbox{.}}{2019}]%
        {dong-etal-2019-document}
\bibfield{author}{\bibinfo{person}{Wei Dong}, \bibinfo{person}{Qinliang Su},
  \bibinfo{person}{Dinghan Shen}, {and} \bibinfo{person}{Changyou Chen}.}
  \bibinfo{year}{2019}\natexlab{}.
\newblock \showarticletitle{Document Hashing with Mixture-Prior Generative
  Models}. In \bibinfo{booktitle}{\emph{EMNLP}}. \bibinfo{pages}{5226--5235}.
\newblock


\bibitem[\protect\citeauthoryear{Greene, Parnas, and Yao}{Greene
  et~al\mbox{.}}{1994}]%
        {greene1994multi}
\bibfield{author}{\bibinfo{person}{Dan Greene}, \bibinfo{person}{Michal
  Parnas}, {and} \bibinfo{person}{Frances Yao}.}
  \bibinfo{year}{1994}\natexlab{}.
\newblock \showarticletitle{Multi-index hashing for information retrieval}. In
  \bibinfo{booktitle}{\emph{Proceedings 35th Annual Symposium on Foundations of
  Computer Science}}. IEEE, \bibinfo{pages}{722--731}.
\newblock


\bibitem[\protect\citeauthoryear{Hansen, Hansen, Simonsen, Alstrup, and
  Lioma}{Hansen et~al\mbox{.}}{2019}]%
        {hansensemhash2019}
\bibfield{author}{\bibinfo{person}{Casper Hansen}, \bibinfo{person}{Christian
  Hansen}, \bibinfo{person}{Jakob~Grue Simonsen}, \bibinfo{person}{Stephen
  Alstrup}, {and} \bibinfo{person}{Christina Lioma}.}
  \bibinfo{year}{2019}\natexlab{}.
\newblock \showarticletitle{Unsupervised Neural Generative Semantic Hashing}.
  In \bibinfo{booktitle}{\emph{SIGIR}}. \bibinfo{pages}{735--744}.
\newblock


\bibitem[\protect\citeauthoryear{Hansen, Hansen, Simonsen, Alstrup, and
  Lioma}{Hansen et~al\mbox{.}}{2020a}]%
        {hansen-coldstart-hash-2020}
\bibfield{author}{\bibinfo{person}{Casper Hansen}, \bibinfo{person}{Christian
  Hansen}, \bibinfo{person}{Jakob~Grue Simonsen}, \bibinfo{person}{Stephen
  Alstrup}, {and} \bibinfo{person}{Christina Lioma}.}
  \bibinfo{year}{2020}\natexlab{a}.
\newblock \showarticletitle{Content-aware Neural Hashing for Cold-start
  Recommendation}. In \bibinfo{booktitle}{\emph{SIGIR}}.
  \bibinfo{pages}{971–980}.
\newblock


\bibitem[\protect\citeauthoryear{Hansen, Hansen, Simonsen, Alstrup, and
  Lioma}{Hansen et~al\mbox{.}}{2020b}]%
        {hansen2020PairRec}
\bibfield{author}{\bibinfo{person}{Casper Hansen}, \bibinfo{person}{Christian
  Hansen}, \bibinfo{person}{Jakob~Grue Simonsen}, \bibinfo{person}{Stephen
  Alstrup}, {and} \bibinfo{person}{Christina Lioma}.}
  \bibinfo{year}{2020}\natexlab{b}.
\newblock \showarticletitle{Unsupervised Semantic Hashing with Pairwise
  Reconstruction}. In \bibinfo{booktitle}{\emph{SIGIR}}.
  \bibinfo{pages}{2009–2012}.
\newblock


\bibitem[\protect\citeauthoryear{Hansen, Hansen, Simonsen, and Lioma}{Hansen
  et~al\mbox{.}}{2021}]%
        {hansen2021hammingDisim}
\bibfield{author}{\bibinfo{person}{Christian Hansen}, \bibinfo{person}{Casper
  Hansen}, \bibinfo{person}{Jakob~Grue Simonsen}, {and}
  \bibinfo{person}{Christina Lioma}.} \bibinfo{year}{2021}\natexlab{}.
\newblock \showarticletitle{Projected Hamming Dissimilarity for Bit-Level
  Importance Coding in Collaborative Filtering}. In
  \bibinfo{booktitle}{\emph{Proceedings of The Web Conference 2021}}.
  \bibinfo{pages}{In print}.
\newblock


\bibitem[\protect\citeauthoryear{Kang and McAuley}{Kang and McAuley}{2019}]%
        {kang2019candidate}
\bibfield{author}{\bibinfo{person}{Wang-Cheng Kang} {and}
  \bibinfo{person}{Julian McAuley}.} \bibinfo{year}{2019}\natexlab{}.
\newblock \showarticletitle{Candidate Generation with Binary Codes for
  Large-Scale Top-N Recommendation}. In \bibinfo{booktitle}{\emph{Proceedings
  of the 28th ACM International Conference on Information and Knowledge
  Management}}. \bibinfo{pages}{1523--1532}.
\newblock


\bibitem[\protect\citeauthoryear{Kingma and Ba}{Kingma and Ba}{2014}]%
        {kingma2014adam}
\bibfield{author}{\bibinfo{person}{Diederik~P Kingma} {and}
  \bibinfo{person}{Jimmy Ba}.} \bibinfo{year}{2014}\natexlab{}.
\newblock \showarticletitle{Adam: A method for stochastic optimization}.
\newblock \bibinfo{journal}{\emph{ICLR}}.
\newblock


\bibitem[\protect\citeauthoryear{Kingma and Welling}{Kingma and
  Welling}{2014}]%
        {kingma2013auto}
\bibfield{author}{\bibinfo{person}{Diederik~P Kingma} {and}
  \bibinfo{person}{Max Welling}.} \bibinfo{year}{2014}\natexlab{}.
\newblock \showarticletitle{Auto-encoding variational bayes}. In
  \bibinfo{booktitle}{\emph{ICLR}}.
\newblock


\bibitem[\protect\citeauthoryear{Leskovec, Rajaraman, and Ullman}{Leskovec
  et~al\mbox{.}}{2020}]%
        {leskovec2020mining}
\bibfield{author}{\bibinfo{person}{Jure Leskovec}, \bibinfo{person}{Anand
  Rajaraman}, {and} \bibinfo{person}{Jeffrey~David Ullman}.}
  \bibinfo{year}{2020}\natexlab{}.
\newblock \bibinfo{booktitle}{\emph{Mining of massive data sets}}.
\newblock \bibinfo{publisher}{Cambridge university press}.
\newblock


\bibitem[\protect\citeauthoryear{Lian, Xie, and Chen}{Lian
  et~al\mbox{.}}{2019}]%
        {lian2019discrete}
\bibfield{author}{\bibinfo{person}{Defu Lian}, \bibinfo{person}{Xing Xie},
  {and} \bibinfo{person}{Enhong Chen}.} \bibinfo{year}{2019}\natexlab{}.
\newblock \showarticletitle{Discrete matrix factorization and extension for
  fast item recommendation}.
\newblock \bibinfo{journal}{\emph{IEEE Transactions on Knowledge and Data
  Engineering}} (\bibinfo{year}{2019}).
\newblock


\bibitem[\protect\citeauthoryear{Liu, Wang, Kumar, and Chang}{Liu
  et~al\mbox{.}}{2011}]%
        {liu2011hashing}
\bibfield{author}{\bibinfo{person}{Wei Liu}, \bibinfo{person}{Jun Wang},
  \bibinfo{person}{Sanjiv Kumar}, {and} \bibinfo{person}{Shih-Fu Chang}.}
  \bibinfo{year}{2011}\natexlab{}.
\newblock \showarticletitle{Hashing with graphs}. In
  \bibinfo{booktitle}{\emph{ICML}}.
\newblock


\bibitem[\protect\citeauthoryear{Maaten and Hinton}{Maaten and Hinton}{2008}]%
        {maaten2008visualizing}
\bibfield{author}{\bibinfo{person}{Laurens van~der Maaten} {and}
  \bibinfo{person}{Geoffrey Hinton}.} \bibinfo{year}{2008}\natexlab{}.
\newblock \showarticletitle{Visualizing data using t-SNE}.
\newblock \bibinfo{journal}{\emph{Journal of machine learning research}}
  \bibinfo{volume}{9}, \bibinfo{number}{Nov} (\bibinfo{year}{2008}),
  \bibinfo{pages}{2579--2605}.
\newblock


\bibitem[\protect\citeauthoryear{McSherry and Najork}{McSherry and
  Najork}{2008}]%
        {10.5555/1793274.1793325}
\bibfield{author}{\bibinfo{person}{Frank McSherry} {and} \bibinfo{person}{Marc
  Najork}.} \bibinfo{year}{2008}\natexlab{}.
\newblock \showarticletitle{Computing Information Retrieval Performance
  Measures Efficiently in the Presence of Tied Scores}. In
  \bibinfo{booktitle}{\emph{Proceedings of the IR Research, 30th European
  Conference on Advances in Information Retrieval}}.
  \bibinfo{pages}{414–421}.
\newblock


\bibitem[\protect\citeauthoryear{Ng, Jordan, and Weiss}{Ng
  et~al\mbox{.}}{2002}]%
        {ng2002spectral}
\bibfield{author}{\bibinfo{person}{Andrew~Y Ng}, \bibinfo{person}{Michael~I
  Jordan}, {and} \bibinfo{person}{Yair Weiss}.}
  \bibinfo{year}{2002}\natexlab{}.
\newblock \showarticletitle{On spectral clustering: Analysis and an algorithm}.
  In \bibinfo{booktitle}{\emph{NeurIPS}}. \bibinfo{pages}{849--856}.
\newblock


\bibitem[\protect\citeauthoryear{Norouzi and Fleet}{Norouzi and Fleet}{2011}]%
        {norouzi2011minimal}
\bibfield{author}{\bibinfo{person}{Mohammad Norouzi} {and}
  \bibinfo{person}{David~J Fleet}.} \bibinfo{year}{2011}\natexlab{}.
\newblock \showarticletitle{Minimal loss hashing for compact binary codes}. In
  \bibinfo{booktitle}{\emph{ICML}}.
\newblock


\bibitem[\protect\citeauthoryear{Norouzi, Punjani, and Fleet}{Norouzi
  et~al\mbox{.}}{2012}]%
        {norouzi2012fast-initial-paper}
\bibfield{author}{\bibinfo{person}{Mohammad Norouzi}, \bibinfo{person}{Ali
  Punjani}, {and} \bibinfo{person}{David~J Fleet}.}
  \bibinfo{year}{2012}\natexlab{}.
\newblock \showarticletitle{Fast search in hamming space with multi-index
  hashing}. In \bibinfo{booktitle}{\emph{2012 IEEE conference on computer
  vision and pattern recognition}}. IEEE, \bibinfo{pages}{3108--3115}.
\newblock


\bibitem[\protect\citeauthoryear{Norouzi, Punjani, and Fleet}{Norouzi
  et~al\mbox{.}}{2013}]%
        {norouzi2012fast}
\bibfield{author}{\bibinfo{person}{Mohammad Norouzi}, \bibinfo{person}{Ali
  Punjani}, {and} \bibinfo{person}{David~J Fleet}.}
  \bibinfo{year}{2013}\natexlab{}.
\newblock \showarticletitle{Fast exact search in hamming space with multi-index
  hashing}.
\newblock \bibinfo{journal}{\emph{IEEE transactions on pattern analysis and
  machine intelligence}} \bibinfo{volume}{36}, \bibinfo{number}{6}
  (\bibinfo{year}{2013}), \bibinfo{pages}{1107--1119}.
\newblock


\bibitem[\protect\citeauthoryear{Robertson, Walker, Jones, Hancock-Beaulieu,
  Gatford, et~al\mbox{.}}{Robertson et~al\mbox{.}}{1995}]%
        {robertson1995okapi}
\bibfield{author}{\bibinfo{person}{Stephen~E Robertson}, \bibinfo{person}{Steve
  Walker}, \bibinfo{person}{Susan Jones}, \bibinfo{person}{Micheline~M
  Hancock-Beaulieu}, \bibinfo{person}{Mike Gatford}, {et~al\mbox{.}}}
  \bibinfo{year}{1995}\natexlab{}.
\newblock \showarticletitle{Okapi at TREC-3}.
\newblock \bibinfo{journal}{\emph{Nist Special Publication Sp}}
  \bibinfo{volume}{109} (\bibinfo{year}{1995}), \bibinfo{pages}{109}.
\newblock


\bibitem[\protect\citeauthoryear{Salakhutdinov and Hinton}{Salakhutdinov and
  Hinton}{2009}]%
        {salakhutdinov2009semantic}
\bibfield{author}{\bibinfo{person}{Ruslan Salakhutdinov} {and}
  \bibinfo{person}{Geoffrey Hinton}.} \bibinfo{year}{2009}\natexlab{}.
\newblock \showarticletitle{Semantic hashing}.
\newblock \bibinfo{journal}{\emph{International Journal of Approximate
  Reasoning}} \bibinfo{volume}{50}, \bibinfo{number}{7} (\bibinfo{year}{2009}),
  \bibinfo{pages}{969--978}.
\newblock


\bibitem[\protect\citeauthoryear{Shan, Jiao, Zhu, and Mao}{Shan
  et~al\mbox{.}}{2018}]%
        {shan2018recurrent}
\bibfield{author}{\bibinfo{person}{Ying Shan}, \bibinfo{person}{Jian Jiao},
  \bibinfo{person}{Jie Zhu}, {and} \bibinfo{person}{JC Mao}.}
  \bibinfo{year}{2018}\natexlab{}.
\newblock \showarticletitle{Recurrent binary embedding for gpu-enabled
  exhaustive retrieval from billion-scale semantic vectors}. In
  \bibinfo{booktitle}{\emph{Proceedings of the 24th ACM SIGKDD International
  Conference on Knowledge Discovery \& Data Mining}}.
  \bibinfo{pages}{2170--2179}.
\newblock


\bibitem[\protect\citeauthoryear{Shen, Su, Chapfuwa, Wang, Wang, Henao, and
  Carin}{Shen et~al\mbox{.}}{2018}]%
        {nash2018}
\bibfield{author}{\bibinfo{person}{Dinghan Shen}, \bibinfo{person}{Qinliang
  Su}, \bibinfo{person}{Paidamoyo Chapfuwa}, \bibinfo{person}{Wenlin Wang},
  \bibinfo{person}{Guoyin Wang}, \bibinfo{person}{Ricardo Henao}, {and}
  \bibinfo{person}{Lawrence Carin}.} \bibinfo{year}{2018}\natexlab{}.
\newblock \showarticletitle{NASH: Toward End-to-End Neural Architecture for
  Generative Semantic Hashing}. In \bibinfo{booktitle}{\emph{ACL}}.
  \bibinfo{pages}{2041--2050}.
\newblock


\bibitem[\protect\citeauthoryear{Wang, Liu, Kumar, and Chang}{Wang
  et~al\mbox{.}}{2015}]%
        {wang2015learning}
\bibfield{author}{\bibinfo{person}{Jun Wang}, \bibinfo{person}{Wei Liu},
  \bibinfo{person}{Sanjiv Kumar}, {and} \bibinfo{person}{Shih-Fu Chang}.}
  \bibinfo{year}{2015}\natexlab{}.
\newblock \showarticletitle{Learning to hash for indexing big data—A survey}.
\newblock \bibinfo{journal}{\emph{Proc. IEEE}} \bibinfo{volume}{104},
  \bibinfo{number}{1} (\bibinfo{year}{2015}), \bibinfo{pages}{34--57}.
\newblock


\bibitem[\protect\citeauthoryear{Wang, Zhang, Sebe, and Shen}{Wang
  et~al\mbox{.}}{2018}]%
        {wang2018survey}
\bibfield{author}{\bibinfo{person}{Jingdong Wang}, \bibinfo{person}{Ting
  Zhang}, \bibinfo{person}{Nicu Sebe}, {and} \bibinfo{person}{Heng~Tao Shen}.}
  \bibinfo{year}{2018}\natexlab{}.
\newblock \showarticletitle{A survey on learning to hash}.
\newblock \bibinfo{journal}{\emph{IEEE transactions on pattern analysis and
  machine intelligence}} \bibinfo{volume}{40}, \bibinfo{number}{4}
  (\bibinfo{year}{2018}), \bibinfo{pages}{769--790}.
\newblock


\bibitem[\protect\citeauthoryear{Wang, Shen, Wang, Yao, Jiang, Qi, and
  Chen}{Wang et~al\mbox{.}}{2019}]%
        {wang2019learning}
\bibfield{author}{\bibinfo{person}{Meng Wang}, \bibinfo{person}{Haomin Shen},
  \bibinfo{person}{Sen Wang}, \bibinfo{person}{Lina Yao},
  \bibinfo{person}{Yinlin Jiang}, \bibinfo{person}{Guilin Qi}, {and}
  \bibinfo{person}{Yang Chen}.} \bibinfo{year}{2019}\natexlab{}.
\newblock \showarticletitle{Learning to Hash for Efficient Search Over
  Incomplete Knowledge Graphs}. In \bibinfo{booktitle}{\emph{2019 IEEE
  International Conference on Data Mining (ICDM)}}. IEEE,
  \bibinfo{pages}{1360--1365}.
\newblock


\bibitem[\protect\citeauthoryear{Weiss, Torralba, and Fergus}{Weiss
  et~al\mbox{.}}{2009}]%
        {weiss2009spectral}
\bibfield{author}{\bibinfo{person}{Yair Weiss}, \bibinfo{person}{Antonio
  Torralba}, {and} \bibinfo{person}{Rob Fergus}.}
  \bibinfo{year}{2009}\natexlab{}.
\newblock \showarticletitle{Spectral hashing}. In
  \bibinfo{booktitle}{\emph{NeurIPS}}. \bibinfo{pages}{1753--1760}.
\newblock


\bibitem[\protect\citeauthoryear{Zhang, Wang, Cai, and Lu}{Zhang
  et~al\mbox{.}}{2010a}]%
        {zhang2010laplacian}
\bibfield{author}{\bibinfo{person}{Dell Zhang}, \bibinfo{person}{Jun Wang},
  \bibinfo{person}{Deng Cai}, {and} \bibinfo{person}{Jinsong Lu}.}
  \bibinfo{year}{2010}\natexlab{a}.
\newblock \showarticletitle{Laplacian co-hashing of terms and documents}. In
  \bibinfo{booktitle}{\emph{ECIR}}. Springer, \bibinfo{pages}{577--580}.
\newblock


\bibitem[\protect\citeauthoryear{Zhang, Wang, Cai, and Lu}{Zhang
  et~al\mbox{.}}{2010b}]%
        {zhang2010self}
\bibfield{author}{\bibinfo{person}{Dell Zhang}, \bibinfo{person}{Jun Wang},
  \bibinfo{person}{Deng Cai}, {and} \bibinfo{person}{Jinsong Lu}.}
  \bibinfo{year}{2010}\natexlab{b}.
\newblock \showarticletitle{Self-taught hashing for fast similarity search}. In
  \bibinfo{booktitle}{\emph{SIGIR}}. ACM, \bibinfo{pages}{18--25}.
\newblock


\bibitem[\protect\citeauthoryear{Zhang, Shen, Liu, He, Luan, and Chua}{Zhang
  et~al\mbox{.}}{2016a}]%
        {zhang2016discrete}
\bibfield{author}{\bibinfo{person}{Hanwang Zhang}, \bibinfo{person}{Fumin
  Shen}, \bibinfo{person}{Wei Liu}, \bibinfo{person}{Xiangnan He},
  \bibinfo{person}{Huanbo Luan}, {and} \bibinfo{person}{Tat-Seng Chua}.}
  \bibinfo{year}{2016}\natexlab{a}.
\newblock \showarticletitle{Discrete collaborative filtering}. In
  \bibinfo{booktitle}{\emph{Proceedings of the 39th International ACM SIGIR
  conference on Research and Development in Information Retrieval}}.
  \bibinfo{pages}{325--334}.
\newblock


\bibitem[\protect\citeauthoryear{Zhang, Wang, Hong, and Chua}{Zhang
  et~al\mbox{.}}{2016b}]%
        {zhang2016play}
\bibfield{author}{\bibinfo{person}{Hanwang Zhang}, \bibinfo{person}{Meng Wang},
  \bibinfo{person}{Richang Hong}, {and} \bibinfo{person}{Tat-Seng Chua}.}
  \bibinfo{year}{2016}\natexlab{b}.
\newblock \showarticletitle{Play and rewind: Optimizing binary representations
  of videos by self-supervised temporal hashing}. In
  \bibinfo{booktitle}{\emph{Proceedings of the 24th ACM international
  conference on Multimedia}}. \bibinfo{pages}{781--790}.
\newblock


\end{thebibliography}
\end{document}